\documentstyle[aps,preprint]{revtex}
\newcommand{\f}[2]{\frac{#1}{#2}}
\newcommand{\s}[1]{\sqrt{#1}}
\begin{document}
\title{Determination of SU(6) Clebsch--Gordan Coefficients and\\Baryon
Mass and Electromagnetic Moment Relations}
\preprint{UCSD/PTH 94-17}
\author{Richard F.\ Lebed}
\address{Department of Physics, University of California at San Diego,
La Jolla, California 92093}
\date{November 1, 1994}
\maketitle
\begin{abstract}
	We compute and tabulate the Clesbsch--Gordan coefficients of
the $SU(6) \supset SU(3) \times SU(2)$ product ${\overline{\bf 56}}
\otimes {\bf 56}$, which are relevant to the nonrelativistic spin-flavor
symmetry of the lightest baryons.  Under the assumption that the
largest representation in this product, the ${\bf 2695}$, gives rise
to operators in a chiral expansion that produce numerically small
effects, we obtain a set of relations among the masses of the baryons,
as well as among their magnetic dipole and higher multipole moments.
We compare the mass relations to experiment, and find numerical
predictions for the $\Sigma^0$-$\Lambda$ mass mixing parameter and
eighteen of the twenty-seven magnetic moments in the ${\bf 56}$.
\end{abstract}
\newpage
\section{Introduction}
	A generation ago, during the mid-1960's, the highly successful
$SU(3)$ model of light flavors developed by Gell-Mann and
Ne'eman~\cite{su3} was generalized to include the spin symmetry
$SU(2)$ in an enlarged spin-flavor symmetry group, $SU(6)$~\cite{su6}.
The increased predictive power of $SU(6)$ over independent $SU(3)
\times SU(2)$ symmetries immediately produced a number of intriguing
results for the baryons, most notably the relative closeness of baryon
octet and decuplet masses, the axial current coefficient ratio $F/D =
2/3$, and the famous magnetic moment ratio $\mu_{p} / \mu_{n} = -3/2$,
which is experimentally true to 3\%.

	Yet two problems with the theory ultimately brought about its
demise.  The first was that the mesons did not seem to fit as well as
the baryons into the theory; for example, why are the baryon octet and
decuplet relatively close in mass, whereas the vector mesons are 2--5
times heavier than their pseudoscalar partners?  Clearly $SU(6)$ is
somehow special to the baryons.  The other problem was much more
serious, and in retrospect seems almost obvious: Mixing the compact,
purely internal flavor symmetry with the noncompact Poincar\'{e}
symmetry of spin angular momentum must and did ultimately lead to some
nonsensical results.  Such considerations gave rise to the various
{\it no-go\/} theorems of the late 1960's, culminating in the
celebrated Coleman--Mandula theorem~\cite{CM}, all forbidding such
hybrid symmetries.

	Nevertheless, there still exists the troubling matter of the
$\mu_p / \mu_n$ ratio and other baryonic ``coincidences.''  Why should
such good predictions exist?  Although the no-go theorems tell us that
$SU(6)$ cannot be an exact symmetry of nature, there is nothing
forbidding it from being a very good {\em approximate\/} symmetry.  If
this is the case, we may expect that a true symmetry of the universe
generates predictions which are very similar to those of $SU(6)$.

	A promising candidate for such a symmetry is provided by
large-$N_c$ QCD~\cite{Nc}.  It has recently been shown that the baryon
sector of large-$N_c$ QCD possesses a contracted spin-flavor
symmetry~\cite{GS,DM,DJM} which is similar, but not identical, to the
$SU(6)$ spin-flavor symmetry.  Results obtained from a consistent
expansion in powers of $1/N_c$ allow one to explain certain results of
chiral perturbation theory (which in turn relies on $SU(3)$
symmetry) that are difficult to understand otherwise.  It is a
phenomenological fact that combinations of hadronic fields
transforming under the largest representations of $SU(3)$ or $SU(6)$
tend to give rise to numerically small results, which is the origin of
relations between hadron parameters.  Often, but not always, this can
be explained by the fact that the largest representations are
accompanied by several powers of small chiral symmetry-breaking
factors and are thus suppressed.  In the large-$N_c$ contracted
spin-flavor symmetry, on the other hand, operators transforming under
larger representations are accompanied by more powers of $1/N_c$; thus
we have a well-defined prescription for identifying theoretically
suppressed combinations of baryonic parameters, or in other words,
relations among the baryons.

	It is therefore a highly relevant problem to analyze the group
theory of the large-$N_c$ contracted spin-flavor symmetry in order to
find and test relations among baryon parameters, namely masses,
electromagnetic moments, and eventually decay widths and scattering
amplitudes.  Interesting new results have been obtained in this
theory~\cite{DJM,NewN,JM}, but the full analysis has not yet been
completed.  It it is also important to uncover, as is done in this
work, the analogous relations within the related symmetry of $SU(6)$
for comparison to the large-$N_c$ results.  A detailed comparison of
the relationships between physical quantities ultimately helps us to
determine how accurately each symmetry reflects reality.

	In $SU(6)$ the well-known octet and decuplet of baryons fill a
single irreducible representation, the ${\bf 56}$; thus the operators
we consider, bilinears in the baryon fields, are exactly those within
the product of this representation with its conjugate, and many of
these Clebsch--Gordan coefficients have not been tabulated previously.
Therefore, this project also has intrinsic value in a mathematical
sense.  We provide a relatively simple and convenient method by which
such group-theoretical factors may be generated.  Once this is
accomplished, we possess all possible information leading to relations
among the baryons that depend only on $SU(6)$ symmetry.  We then need
to decide only which product representations may be neglected in order
to obtain the desired relations, and test their validity with
experimental inputs.

	This paper is organized as follows: In Sec.\ 2, we begin with
a discussion of $SU(3)$ and its Clebsch--Gordan coefficients, and how
we may use them to build up the corresponding coefficients for
$SU(6)$.  As a warmup, we review the derivation of $SU(3)$ mass
relations using these coefficients in Sec.\ 3.  We explain in Sec.\ 4
the method of computation of the $SU(6)$ coefficients and their
classification by additional $SU(3)$ and isospin quantum numbers.
Tables of the $SU(6)$ Clebsch--Gordan coefficients, and the means by
which relations are derived, are presented in Sec.\ 5\@.  The baryon
relations for masses and magnetic dipole, electric quadrupole, and
magnetic octupole moments are collected in Sec.\ 6; we then use
experimental values to evaluate these relations wherever possible, and
estimate the size of neglected terms.  We summarize our conclusions in
Sec.\ 7.
\section{$SU(3)$ Structure of the Baryons}
    We begin with a systematic classification of $SU(3)$
representations (hereafter ${\it reps\/}$) of the octet and decuplet
baryon field bilinears.  Consider, within the effective Lagrangian,
any term connecting single initial and final baryons respectively
transforming under $R_1$- and $R_2$-dimensional reps:
\begin{equation}
\delta {\cal L} = \overline{R_2} {\cal O} R_1 ,
\end{equation}
where ${\cal O}$ is some operator.  The pattern of $SU(3)$ breaking by
this term is exhibited by the decomposition of $(\overline{R_2} \times
R_1)$ into combinations transforming under all possible irreducible
reps.  For the octet and decuplet, these reps are
\begin{equation} \label{8x8}
{\bf 8} \otimes {\bf 8} = {\bf 1} \oplus {\bf 8}_1 \oplus {\bf 8}_2 \oplus
{\bf 10} \oplus \overline{\bf 10} \oplus {\bf 27},
\end{equation}
\begin{equation}
{\bf 8} \otimes {\bf 10} = {\bf 8} \oplus {\bf 10} \oplus {\bf 27}
\oplus {\bf 35},
\end{equation}
(and its conjugate form $\overline{\bf 10} \otimes {\bf 8}$), and
\begin{equation}
\overline{\bf 10} \otimes {\bf 10} = {\bf 1} \oplus {\bf 8} \oplus {\bf 27}
\oplus {\bf 64}.
\end{equation}
The projections of ${\cal O}$ forming the coefficients of these
combinations can be labeled with the $SU(3)$ indices of the
corresponding bilinear combinations.  We may then loosely speak of
${\cal O}$ as transforming under some rep, although in fact only the
baryon field bilinears transform.  This analysis is, of course, not
restricted to $SU(3)$; its verity relies only on negligible mixing
from heavier states possessing the same quantum numbers.

    A restriction we now place on the baryon terms in the Lagrangian
is that they originate only in the strong and electromagnetic but not
the weak interactions.  That is, we consider only bilinears that
conserve strangeness as well as electric charge, or equivalently,
those with the properties $\Delta I_3 = 0$ and $\Delta Y = 0$.  Note
that these include ``mixing'' terms for any states with the
same values of $I_3$ and $Y$; every octet state mixes with exactly one
decuplet state, and within the octet, $\Sigma^0$-$\Lambda$ mixing can occur.

    It remains only to distinguish degenerate $\Delta I_3 = \Delta Y =
0$ operators within a rep.  As usual, we assume the standard
notation of labeling with the isospin Casimir $I(I+1)$, so that
$x^R_I$ (where $x$ is a generic coefficient name) specifies a unique
chiral coefficient within the rep ${\bf R}$.  It then becomes a
straightforward exercise with the well-known $SU(3)$ Clebsch--Gordan
coefficients (see, {\it e.g.}, Ref.~\cite{deS}) to decompose bilinear
terms into the forms
\begin{eqnarray*}
{\bf M}_a  & = & {\cal C}_a {\bf a} , \\
{\bf M}_b  & = & {\cal C}_b {\bf b} , \\
{\bf M}_c  & = & {\cal C}_c {\bf c} , \\
{\bf M}_{\overline{c}} & = & {\cal C}_{\overline{c}} {\overline{\bf c}},
\end{eqnarray*}
where
\begin{displaymath}
{\bf M}_a \equiv \left( \! \begin{array}{c}
\overline{p} p               \\ \overline{n} n               \\
\overline{\Sigma^+} \Sigma^+ \\ \overline{\Sigma^0} \Sigma^0 \\
\overline{\Lambda} \Lambda   \\ \overline{\Sigma^-} \Sigma^- \\
\overline{\Xi^0} \Xi^0       \\ \overline{\Xi^-} \Xi^-       \\
\overline{\Sigma^0} \Lambda  \\ \overline{\Lambda} \Sigma^0
\end{array} \! \right) , \hspace{2ex}
{\bf M}_b \equiv \left( \! \begin{array}{c}
\overline{\Delta^{+ \! +}} \Delta^{+ \! +} \\ \overline{\Delta^+} \Delta^+ \\
\overline{\Delta^0} \Delta^0       \\ \overline{\Delta^-} \Delta^-         \\
\overline{\Sigma^{*+}} \Sigma^{*+} \\ \overline{\Sigma^{*0}} \Sigma^{*0}   \\
\overline{\Sigma^{*-}} \Sigma^{*-} \\ \overline{\Xi^{*0}} \Xi^{*0}         \\
\overline{\Xi^{*-}} \Xi^{*-}       \\ \overline{\Omega^-} \Omega^-
\end{array} \! \right) , \hspace{2ex}
{\bf M}_c \equiv \left( \! \begin{array}{c}
\overline{p} \Delta^+           \\ \overline{n} \Delta^0           \\
\overline{\Sigma^+} \Sigma^{*+} \\ \overline{\Sigma^0} \Sigma^{*0} \\
\overline{\Lambda} \Sigma^{*0}  \\ \overline{\Sigma^-} \Sigma^{*-} \\
\overline{\Xi^0} \Xi^{*0}       \\ \overline{\Xi^-} \Xi^{*-}
\end{array} \! \right) , \hspace{2ex}
{\bf M}_{\overline{c}} \equiv \left( \! \begin{array}{c}
\overline{\Delta^+} p           \\ \overline{\Delta^0} n           \\
\overline{\Sigma^{*+}} \Sigma^+ \\ \overline{\Sigma^{*0}} \Sigma^0 \\
\overline{\Sigma^{*0}} \Lambda  \\ \overline{\Sigma^{*-}} \Sigma^- \\
\overline{\Xi^{*0}} \Xi^0       \\ \overline{\Xi^{*-}} \Xi^-
\end{array} \! \right) ,
\end{displaymath}
\vspace{10ex}
\begin{eqnarray}
{\cal C}_a & = &
\left( \begin{array}{cccccccccc}
+\f{1}{2\s{2}} & +\f{1}{2\s{5}}         & -\f{1}{2}\s{\f{3}{5}} &
+\f{1}{2}      & +\f{1}{2\s{3}}         & +\f{1}{2\s{3}}        &
+\f{1}{2\s{3}} & +\f{1}{2}\s{\f{3}{10}} & +\f{1}{\s{10}}        & 0 \\
+\f{1}{2\s{2}} & +\f{1}{2\s{5}}         & +\f{1}{2}\s{\f{3}{5}} &
+\f{1}{2}      & -\f{1}{2\s{3}}         & -\f{1}{2\s{3}}        &
-\f{1}{2\s{3}} & +\f{1}{2}\s{\f{3}{10}} & -\f{1}{\s{10}}        & 0 \\
+\f{1}{2\s{2}} & -\f{1}{\s{5}}  & 0              & 0               &
+\f{1}{\s{3}}  & -\f{1}{2\s{3}} & -\f{1}{2\s{3}} & -\f{1}{2\s{30}} &
0              & +\f{1}{\s{6}}  \\
+\f{1}{2\s{2}} & -\f{1}{\s{5}} & 0 & 0 & 0 & 0 & 0 & -\f{1}{2\s{30}} & 0 &
 -\s{\f{2}{3}} \\
+\f{1}{2\s{2}} & +\f{1}{\s{5}} & 0 & 0 & 0 & 0 & 0 & -\f{3}{2}\s{\f{3}{10}} &
0              & 0             \\
+\f{1}{2\s{2}} & -\f{1}{\s{5}}  & 0              & 0               &
-\f{1}{\s{3}}  & +\f{1}{2\s{3}} & +\f{1}{2\s{3}} & -\f{1}{2\s{30}} &
0              & +\f{1}{\s{6}}  \\
+\f{1}{2\s{2}} & +\f{1}{2\s{5}}         & +\f{1}{2}\s{\f{3}{5}} &
-\f{1}{2}      & +\f{1}{2\s{3}}         & +\f{1}{2\s{3}}        &
+\f{1}{2\s{3}} & +\f{1}{2}\s{\f{3}{10}} & -\f{1}{\s{10}}        & 0 \\
+\f{1}{2\s{2}} & +\f{1}{2\s{5}}         & -\f{1}{2}\s{\f{3}{5}} &
-\f{1}{2}      & -\f{1}{2\s{3}}         & -\f{1}{2\s{3}}        &
-\f{1}{2\s{3}} & +\f{1}{2}\s{\f{3}{10}} & +\f{1}{\s{10}}        & 0 \\
 0 & 0 & +\f{1}{\s{5}} & 0 & 0 & +\f{1}{2} & -\f{1}{2} & 0 & +\s{\f{3}{10}} &
0 \\
 0 & 0 & +\f{1}{\s{5}} & 0 & 0 & -\f{1}{2} & +\f{1}{2} & 0 & +\s{\f{3}{10}} &
0 \\
\end{array} \right) , \nonumber
\end{eqnarray}
\vspace{0.5in}
\begin{eqnarray}
{\cal C}_b = & &
\left( \begin{array}{cccccccccc}
+\f{1}{\s{10}} & +\f{1}{\s{10}} & +\s{\f{3}{10}}   & +\s{\f{3}{70}}          &
+\f{3}{\s{70}} & +\s{\f{3}{14}} & +\f{1}{2 \s{35}} & +\f{1}{2} \s{\f{3}{35}} &
+\f{1}{2\s{7}} & +\f{1}{2\s{5}} \\
+\f{1}{\s{10}} & +\f{1}{\s{10}} & +\f{1}{\s{30}}  & +\s{\f{3}{70}}   &
+\f{1}{\s{70}} & -\s{\f{3}{14}} & +\f{1}{2\s{35}} & +\f{1}{2\s{105}} &
-\f{1}{2\s{7}} & -\f{3}{2\s{5}} \\
+\f{1}{\s{10}} & +\f{1}{\s{10}} & -\f{1}{\s{30}}  & +\s{\f{3}{70}}   &
-\f{1}{\s{70}} & -\s{\f{3}{14}} & +\f{1}{2\s{35}} & -\f{1}{2\s{105}} &
-\f{1}{2\s{7}} & +\f{3}{2\s{5}} \\
+\f{1}{\s{10}} & +\f{1}{\s{10}} & -\s{\f{3}{10}}  & +\s{\f{3}{70}}         &
-\f{3}{\s{70}} & +\s{\f{3}{14}} & +\f{1}{2\s{35}} & -\f{1}{2}\s{\f{3}{35}} &
+\f{1}{2\s{7}} & -\f{1}{2\s{5}} \\
+\f{1}{\s{10}} & 0              & +\s{\f{2}{15}} & -\s{\f{5}{42}} &
-\f{3}{\s{70}} & +\f{1}{\s{42}} & -\f{2}{\s{35}} & -\s{\f{5}{21}} &
-\f{1}{\s{7}}  & 0              \\
+\f{1}{\s{10}} & 0 & 0             & -\s{\f{5}{42}} & 0 & -\s{\f{2}{21}} &
-\f{2}{\s{35}} & 0 & +\f{2}{\s{7}} & 0 \\
+\f{1}{\s{10}} & 0              & -\s{\f{2}{15}} & -\s{\f{5}{42}} &
+\f{3}{\s{70}} & +\f{1}{\s{42}} & -\f{2}{\s{35}} & +\s{\f{5}{21}} &
-\f{1}{\s{7}}  & 0 \\
+\f{1}{\s{10}}      & -\f{1}{\s{10}} & +\f{1}{\s{30}} & -\s{\f{3}{70}} &
-\f{2\s{2}}{\s{35}} & 0              & +\f{3}{\s{35}} & +\s{\f{5}{21}} &
0                   & 0              \\
+\f{1}{\s{10}}      & -\f{1}{\s{10}} & -\f{1}{\s{30}} & -\s{\f{3}{70}} &
+\f{2\s{2}}{\s{35}} & 0              & +\f{3}{\s{35}} & -\s{\f{5}{21}} &
0                   & 0              \\
+\f{1}{\s{10}} & -\s{\f{2}{5}} & 0 & +\f{3\s{3}}{\s{70}} & 0 & 0 &
-\f{2}{\s{35}} & 0 & 0 & 0 \\
\end{array} \right) , \nonumber
\end{eqnarray}
\vspace{10ex}
\begin{eqnarray}
{\cal C}_c = & &
\left( \begin{array}{cccccccc}
 0              & +\f{2}{\s{15}}     & +\f{1}{\s{6}}  & 0              &
+\f{1}{2\s{10}} & +\f{\s{3}}{2\s{2}} & +\f{1}{2\s{6}} & +\f{1}{2\s{2}} \\
 0              & +\f{2}{\s{15}}     & +\f{1}{\s{6}}  & 0              &
+\f{1}{2\s{10}} & -\f{\s{3}}{2\s{2}} & +\f{1}{2\s{6}} & -\f{1}{2\s{2}} \\
+\f{1}{\s{5}}   & +\f{1}{\s{15}}     & -\f{1}{\s{6}}  & +\s{\f{2}{15}} &
+\f{3}{2\s{10}} & +\f{1}{2\s{6}}     & -\f{1}{2\s{6}} & -\f{1}{2\s{2}} \\
+\f{1}{\s{5}}   & 0                  & 0              & +\s{\f{2}{15}} &
 0              & -\f{1}{\s{6}}      & 0              & +\f{1}{\s{2}}  \\
 0              & +\f{1}{\s{5}}      & 0              & 0              &
-\s{\f{3}{10}}  & 0                  & -\f{1}{\s{2}}  & 0              \\
+\f{1}{\s{5}}   & -\f{1}{\s{15}}     & +\f{1}{\s{6}}  & +\s{\f{2}{15}} &
-\f{3}{2\s{10}} & +\f{1}{2\s{6}}     & +\f{1}{2\s{6}} & -\f{1}{2\s{2}} \\
+\f{1}{\s{5}}   & +\f{1}{\s{15}}     & -\f{1}{\s{6}}  & -\s{\f{3}{10}} &
-\f{1}{\s{10}}  & 0                  & +\f{1}{\s{6}}  & 0              \\
+\f{1}{\s{5}}   & -\f{1}{\s{15}}     & +\f{1}{\s{6}}  & -\s{\f{3}{10}} &
+\f{1}{\s{10}}  & 0                  & -\f{1}{\s{6}}  & 0              \\
\end{array} \right) , \nonumber
\end{eqnarray}
\vspace{5ex} \\
with ${\cal C}_{\overline{c}} = {\cal C}_c$, and
\begin{equation} \label{M}
{\bf a} \equiv \left( \begin{array}{c}
a^1_0 \\ a^{8_1}_0 \\ a^{8_1}_1 \\ a^{8_2}_0 \\ a^{8_2}_1 \\ a^{10}_1 \\
a^{\overline{10}}_1 \\ a^{27}_0 \\ a^{27}_1 \\ a^{27}_2
\end{array} \right) , \hspace{2ex}
{\bf b} \equiv \left( \begin{array}{c}
b^1_0 \\ b^8_0 \\ b^8_1 \\ b^{27}_0 \\ b^{27}_1 \\ b^{27}_2 \\ b^{64}_0 \\
b^{64}_1 \\ b^{64}_2 \\ b^{64}_3 \end{array} \right) , \hspace{2ex}
{\bf c} \equiv \left( \begin{array}{c}
c^8_0 \\ c^8_1 \\ c^{10}_1 \\ c^{27}_0 \\ c^{27}_1 \\ c^{27}_2 \\
c^{35}_1 \\ c^{35}_2
\end{array} \right), \hspace{2ex}
\overline{\bf c} \equiv \left( \begin{array}{c}
\overline{c}^8_0 \\ \overline{c}^8_1 \\ \overline{c}^{\overline{10}}_1
\\ \overline{c}^{27}_0 \\ \overline{c}^{27}_1 \\ \overline{c}^{27}_2
\\ \overline{c}^{\overline{35}}_1 \\ \overline{c}^{\overline{35}}_2
\end{array} \right) .
\end{equation}

    Here the ${\bf 8} \otimes {\bf 8}$ reps ${\bf 8}_{1,2}$ are
distinguished by the symmetry properties of their components under
reflection through the origin in weight space ({\it i.e.}, exchanging
the component transforming with quantum numbers ($I$,$I_3$,$Y$) with
that transforming under ($I$,$-I_3$,$-Y$)).  ${\bf 8}_{1,2}$ is
symmetric (antisymmetric) under this exchange, giving, for instance,
the same (opposite) contributions to the bilinears of the $p$ and
$\Xi^-$.

    With the above normalization of the chiral coefficients
$a$, $b$, $c$, and $\overline{c}$, the matrices ${\cal C}$ are
orthogonal.  This, of course, must be the the case, for we are merely
describing the bilinears in a different basis.  Because the matrices
are orthogonal, we may alter the sign of any row or column and still
maintain orthogonality.  The phase conventions exhibited above have
been chosen ultimately to match well-known quark-model results; for
example, each octet term has the same singlet coefficient $a^1_0 \, /
2\s{2}$.  We are thus fixing the phases of the {\em lowest}-weight
reps, the direct opposite of the usual Condon--Shortley convention.

    It is easy to understand the number of chiral coefficients
appearing in the octet and decuplet products.  With arbitrary $SU(3)$
breaking, one may clearly supply each bilinear with a distinct
arbitrary coefficient; hence the decuplet product must have ten chiral
coefficients, the decuplet-octet product eight, and the octet product
ten, because the octet supports $\Sigma^0$-$\Lambda$ mixing.  But such
mixing requires only one parameter, a mixing angle $\theta$.  In the
above matrices there are two, corresponding to the bilinears
$\overline{\Sigma^0} \Lambda$ and $\overline{\Lambda}
\Sigma^0$.  However, hermiticity (or time-reversal invariance) of the
Lagrangian reduces these to one, imposing the physical constraint
$a^{10}_1 = -a^{\overline{10}}_1$.  Later we find a similar constraint
between $c^R_I$ and $\overline{c}^{\overline{R}}_I$.

	Complete knowledge of the $SU(3)$ group-theoretical factors
already tells us a great deal about the corresponding factors for
$SU(6)$, for the quantum numbers of the latter symmetry group are
assigned via the decomposition $SU(6) \supset SU(3) \times SU(2)$, and
the flavor and spin groups commute.  Thus a chiral coefficient of any
rep ${\bf N}$ of $SU(6)$, distinguished by its decomposition into an
$R$-dimensional rep of $SU(3)$ and isospin $I$ (henceforth denoted by
$d_N^{R,I}$) must be some linear combination of all existing chiral
coefficients $a^R_I$, $b^R_I$, $c^R_I$, and $\overline{c}^R_I$.  For
example, because spin and flavor commute, the bilinear combination
$a^{8_1}_1$ still tranforms as the $I=1$ component of an octet
regardless of how we insert spins on the baryon indices.  Thus, since
the combinations $a^{8_1}_1$, $a^{8_2}_1$, $b^8_1$, $c^8_1$, and
$\overline{c}^8_1$ span the entire subspace of $I=1$ octets formed
from the baryon octet and decuplet bilinears, each $d^{8,1}_N$ must be
a linear combination of these.
\section{Example: $SU(3)$ Baryon Mass Relations}
	As a preliminary to $SU(6)$, let us consider how to obtain
relations between the baryons using only $SU(3)$ group theory.
Because the latter multiplets take into account only flavor symmetry,
we do not expect to learn anything about quantities in which the
individual spin states are important (${\it e.g.\/}$ magnetic moment
relations).  However, we can learn about the masses.  First we assume
that mixing between multiplets is negligible, so that the physical
baryons truly live in octet and decuplet reps of $SU(3)$.  In the
usual chiral Lagrangian, $SU(3)$ breaking is accomplished by an
expansion in the quark mass ($M_q$) and charge ($Q_q$) operators; in
terms of flavor indices, these are $3 \times 3$ matrices (with
$u$,$d$, and $s$ diagonal entries), and such operators $X$ may be
decomposed into octet $(X-\f{1}{3}({\rm Tr}X) {\bf 1})$ and singlet
$(({\rm Tr}X) {\bf 1})$ portions.  At first order in $SU(3)$ breaking
only these singlet and octet operators are present; at second order,
operators in the reps of Eq.~\ref{8x8} appear.

	An important reason for the success of the chiral Lagrangian
formalism is that the operators $M_q$ and $Q_q$ enter into the
Lagrangian as operators with perturbatively small coefficients; in the
case of $M_q$, contributions are suppressed by at least $m_s /
\Lambda_{\chi} \approx 0.2$, where $m_s$ is the strange quark mass,
and $\Lambda_{\chi}$ is the chiral symmetry-breaking scale.  Terms
involving only $m_{u,d}$ are suppressed by another factor of 20 or so.
For $Q_q$, the suppression comes through powers of $e \approx 0.3$,
although in mass relations, charge conjugation symmetry of the strong
and electromagnetic interactions permits factors of $Q_q$ only in even
numbers; there is a further suppressions of $16\pi^2$ because such
mass terms come from photon loop effects in the QCD Lagrangian.  Thus
the true suppression is by $\alpha/4\pi \approx 6 \times 10^{-4}$.

	So now we can see explicitly why the coefficients associated
with the largest reps are suppressed: Larger reps require more powers
of the small symmetry-breaking reps, which in turn bring in more
numerical suppressions.

	Let us consider some examples, first supposing that splittings
within i\-so\-spin multiplets are negligible.  Then all chiral
coefficients of the form $x^R_I$ with $I>0$ must also be negligible.
In this case, the only independent octet masses are $N$, $\Sigma$,
$\Lambda$, and $\Xi$, whereas the only nontrivial chiral coefficients
are $a^1_0$, $a^{8_1}_0$, $a^{8_2}_0$, and $a^{27}_0$.  If we only
work to first order in $SU(3)$ breaking, the last of these is
identically zero, and we find
\begin{equation} \label{gmo}
\begin{array}{rcccccl}
\Delta_{\rm GMO} & \equiv & \f{1}{2} \s{\f{10}{3}} a^{27}_0 & = & \f{3}{4}
\Lambda + \f{1}{4} \Sigma -\f{1}{2} (N + \Xi) & = & 0,
\end{array}
\end{equation}
the Gell-Mann--Okubo relation~\cite{GMO}.  For the decuplet, the
independent masses are $\Delta$, $\Sigma^*$, $\Xi^*$, and $\Omega$,
whereas the nontrivial chiral coefficients are $b^1_0$, $b^8_0$,
$b^{27}_0$, and $b^{64}_0$.  To first order in $SU(3)$ breaking, the
vanishing of the last two coefficients gives rise to two nontrivial
relations, which may be written
\begin{equation} \label{esr}
\begin{array}{rcrcccc}
0 & = & 5(2b^{27}_0 + b^{64}_0) & = & (\Delta - \Sigma^*) & - &
(\Sigma^* - \Xi^*), \\
0 & = & 10(b^{27}_0 - 2b^{64}_0) & = & (\Sigma^* - \Xi^*) & - &
(\Omega - \Xi^*) ,
\end{array}
\end{equation}
Gell-Mann's famous equal-spacing rule~\cite{GlM}.

	On the other hand, if we consider only $I=2$ operators (which
we expect to be numerically well-suppressed by $\alpha /4\pi$ or
$(m_u-m_d)^2/\Lambda_{\chi}^2)$, the octet provides us with the
$\Sigma$ equal-spacing rule~\cite{CG}:
\begin{equation} \label{sig}
\begin{array}{rcccccc}
\Delta_{\Sigma} & \equiv & \s{6} a^{27}_2 & = &
(\Sigma^+ - \Sigma^0 ) & - & (\Sigma^0 - \Sigma^- ) .
\end{array}
\end{equation}
We caution that $\Sigma^0$ in this equation refers to the isospin
$I=1$ eigenstate rather than the mass eigenstate.  In fact, we display
in Sec.\ 6 a new $SU(6)$ relation for the mixing parameter.

	Now consider second-order terms in $SU(3)$ breaking.  {\it A
priori\/} we might expect to find that all of the representations
within the product ${\bf 8} \otimes {\bf 8}$ occur, but we show that
this is not the case.  Because of charge conjugation symmetry of the
strong interaction, the mass Lagrangian contains no terms with an odd
number of $Q_q$ factors.  Thus the only second-order terms in $SU(3)$
breaking are of the forms $(M_q \times M_q)$ and $(Q_q \times Q_q)$.
Consider the product of two identical arbitrary matrices: $(X \times
X)_{ij}{}^{kl}$, which contains such terms as $X_i{}^k X_j{}^l$,
$X_i{}^l X_j{}^k$, and various traces of $X$, where $i,j,k,l$ are
flavor indices in the usual notation.  It is readily seen that this
product has no piece transforming under a {\bf 10}, for such a tensor
with the given indices has the form $A_{ijm} \epsilon^{mkl}$, and is
symmetric under permutation of $\{i,j,m\}$.  If we attempt to
construct a product with these symmetry properties from two identical
matrices, we quickly see that such a term vanishes.  Similarly, the
product of two identical matrices may contain no piece of a
$\overline{\bf 10}$.

    We conclude that, to second order in $SU(3)$ breaking, the octet chiral
coefficients $a^{10}_1 = a^{\overline{10}}_1$ are zero.  The baryon mass
relation corresponding to the vanishing of these coefficients is
\begin{equation} \label{cg}
\Delta_{\rm CG} \equiv -2 \s{3} a^{10}_1 = (n-p) + (\Sigma^+ - \Sigma^-) -
(\Xi^0 - \Xi^-) = 0 ,
\end{equation}
the Coleman--Glashow relation~\cite{CG}.  For the decuplet, the
analysis is even easier: ${\bf 8} \otimes {\bf 8}$ contains no {\bf 64}
for {\em arbitrary\/} pairs of $3 \times 3$ matrices, and so we have
four mass relations good to second-order in $SU(3)$ breaking,
corresponding to the vanishing of $b^{64}_{0,1,2,3}$:
\begin{eqnarray}
\hspace{-4.5em} \Delta_1 \equiv \hspace{3.21em} 20 b^{64}_3 & = & \Delta^{++} -
3\Delta^{+} + 3 \Delta^{0} - \Delta^{-} , \label{first} \\
\hspace{-4.5em} \Delta_2 \equiv \hspace{3.21em} 28 b^{64}_2 & = & \left(
\Delta^{++} - \Delta^{+} - \Delta^{0} + \Delta^{-} \right) - 2 \left(
\Sigma^{*+} - 2 \Sigma^{*0} + \Sigma^{*-} \right) , \\
\hspace{-2em} \Delta_3 \equiv 6(7b^{64}_1 - b^{64}_3) & = & \left( \Delta^{+}
- \Delta^{0} \right) - \left( \Sigma^{*+} - \Sigma^{*-} \right) + \left(
\Xi^{*0} - \Xi^{*-} \right), \\
\hspace{-4.5em} \Delta_4 \equiv \hspace{3.21em} 35 b^{64}_0 & = & \frac{1}{4}
\left( \Delta^{++} + \Delta^{+} + \Delta^{0} + \Delta^{-} \right) - \left(
\Sigma^{*+} + \Sigma^{*0} + \Sigma^{*-} \right) \label{3rd} \nonumber \\
& & \mbox{} + \frac{3}{2} \left( \Xi^{*0} + \Xi^{*-}
\right) - \Omega^{-} \label{last}
\end{eqnarray}
are four vanishing combinations.  Notice that the first three of these
are isospin-breaking, and only the fourth remains in the limit that
isospin is a good symmetry.  The Gell-Mann--Okubo, Coleman--Glashow,
and $\Sigma$ equal-spacing relations and their violations were
explored in chiral perturbation theory in Ref.~\cite{L1}, whereas
similar computations for the relations Eqs.~\ref{first}--\ref{last}
were performed in Ref.~\cite{L2}.

	The approach of identifying relations with large, highly
suppressed reps of course applies to any symmetry group, and we now
proceed to apply it to $SU(6)$.  First, however, we must generate the
orthogonal matrix of spin-flavor baryon bilinears analogous to those
in Eq.~\ref{M}.
\section{Determination of SU(6) Clebsch-Gordan Coefficients}
	The orthogonal matrix of $SU(6)$ group-theoretical factors can
be determined most easily using tensor methods, in a manner similar to
that in which we identified $SU(3)$ mass relations in the previous
Section.  In this case the basic reps in $SU(6)$ breaking are no
longer octets, but $6 \times 6$ traceless matrices, the ${\bf 35}$
(adjoint) rep.  The spin-1/2 octet (16 states) and spin-3/2 decuplet
(40 states) of baryons neatly fill out the ${\bf 56}$ rep, and thus
the relevant products for our analysis are
\begin{equation}
\overline{\bf 56} \otimes {\bf 56} = {\bf 1} \oplus {\bf 35} \oplus
{\bf 405} \oplus {\bf 2695}
\end{equation}
and
\begin{equation}
{\bf 35} \otimes {\bf 35} = {\bf 1} \oplus {\bf 35}_1 \oplus {\bf
35}_2 \oplus {\bf 189} \oplus {\bf 280} \oplus \overline{\bf 280}
\oplus {\bf 405}.
\end{equation}
In particular, since the ${\bf 2695}$ rep does not occur in the latter
product, combinations transforming under this rep give rise to
relations broken only at third order.

	The most straightforward approach to computing the necessary
coefficients is to use the standard Wigner method of starting with the
highest-weight state of the $\overline{\bf 56} \otimes {\bf 56}$
product (which is $\overline{\Delta^-_{-\f{3}{2}}} \Delta^{++}_
{+\f{3}{2}}$) and applying successive $SU(6)$ lowering operators,
orthogonalizing degenerate states as necessary.  Such an approach
gives us not only the $\Delta I_3 = \Delta Y = 0$ bilinears, but all
$56^2 = 3132$ of them.

	This is vastly more effort than we need to expend.  To
demonstrate the point, let us perform a counting of the bilinears we
need: In addition to $\Delta I_3 = \Delta Y = 0$, we also impose
$\Delta J_3 = 0$, where $J$ is total spin of the bilinear.  Using
again that spin and flavor commute in $SU(6)$, we can obtain any
$\Delta J_3 \neq 0$ by means of the simple $SU(2)$ Wigner-Eckart
theorem.  Because the octet is spin-1/2 and the decuplet spin-3/2,
octet-octet bilinears may appear only with $J=0,1$, octet-decuplet
with $J=1,2$, and decuplet-decuplet with $J=0,1,2,3$, and each $J$
multiplet possesses a unique $J_3 = 0$ state.  Recalling from the
previous Section that the number of independent flavor bilinears (not
counting hermiticity) in the ${\bf 8} \otimes {\bf 8}$, ${\bf 8}
\otimes {\bf 10}$, $\overline{\bf 10} \otimes {\bf 8}$, and
$\overline{\bf 10} \otimes {\bf 10}$ products are 10, 8, 8, and 10
respectively, we find
\begin{displaymath}
10(1+1) + 8(1+1) + 8(1+1) + 10(1+1+1+1) = 92
\end{displaymath}
independent baryon bilinears with $\Delta I_3 = \Delta Y = \Delta J_3
= 0$.  The central thrust of this paper, therefore, is the computation
of a $92 \times 92$ orthogonal matrix.

	In fact this task is simplified by the observation that the
combinations of physical relevance are actually those with a
well-defined $J$ quantum number: $J=0$ provides us with information
about the baryon masses (also their ``electric monopole moments,'' or
charges, although this information is of course trivial), $J=1$ tells
us about their magnetic dipole moments, and $J=2,3$ about their
electric quadrupole and magnetic octupole moments, respectively.  This
approach block-diagonalizes the $92 \times 92$ matrix according to
values of $J$.  Performing the counting above including only the
single $J_3$ operator relevant to each value of $J$, we find that the
$J=0,1,2,3$ blocks are respectively square matrices with 20, 36, 26,
and 10 elements on a side.  This is certainly a far cry from the full
matrix of {\em all\/} bilinears, which has $56^2$ entries---on each
side!

	There are yet further simplifications to this approach.  Many
of the entries will be related by means of hermiticity of the
Lagrangian.  We have seen already in $SU(3)$ how this relates the two
$\Sigma^0$-$\Lambda$ bilinears; the same must be true for bilinears
like $\overline{p} \Delta^+$ and $\overline{\Delta^+} p$.  Consequently,
the chiral coefficients of octet-decuplet mixing appear only in certain
characteristic combinations.  We find that, of the 92 parameters at
our disposal, the hermiticity constraint reduces this number to 74.

	The next task is to find the $SU(3) \times SU(2)$ content of
the $SU(6)$ multiplets.  This can be accomplished by forming the
products of the Young tableaux for $SU(3)$ and $SU(2)$ in parallel
with those for $SU(6)$, adding one block ({\it i.e.\/} fundamental rep
index) at a time for each symmetry group.  Then the content of an
$SU(6)$ rep must be such that the sum of the products of $SU(3)$ and
$SU(2)$-rep multiplicities adds up to the multiplicity of the $SU(6)$
rep.  As a simple example, in $SU(6)$ the product of fundamental
conjugate and fundamental reps is
\begin{equation}
\overline{\bf 6} \otimes {\bf 6} = {\bf 1} \oplus {\bf 35},
\end{equation}
whereas for $SU(3)$ and $SU(2)$ the corresponding products are
\begin{eqnarray}
\overline{\bf 3} \otimes {\bf 3} & =  & {\bf 1} \oplus {\bf 8}, \\
{\bf 2} \otimes {\bf 2} & = & {\bf 1} \oplus {\bf 3}.
\end{eqnarray}
So writing $SU(3) \times SU(2)$ content reps as ($R$, $2I+1$), we have
\begin{equation}
{\bf 1} = (1,1) ; \hspace{2em} {\bf 35} = (1,3) + (8,1) + (8,3).
\end{equation}
As long as we construct products one fundamental index at a time, there
is never an ambiguity about how to assign content reps (at least
for the $\overline{\bf 56} \otimes {\bf 56}$ product).  We find
the following decomposition for each value of J:
\begin{eqnarray}
& \underline{J=0} & \nonumber \\
\underline{SU(6) \mbox{ rep}} & & \underline{SU(3) \mbox{ content reps}}
\nonumber \\
{\bf 1}    & & {\bf 1} \nonumber \\
{\bf 35}   & & {\bf 8} \nonumber \\
{\bf 405}  & & {\bf 1}, {\bf 8}, {\bf 27} \nonumber \\
{\bf 2695} & & {\bf 8}, {\bf 10}, \overline{\bf 10}, {\bf 27}, {\bf
64} \hspace{2em}, \\ & & \nonumber \\
& \underline{J=1} & \nonumber \\
\underline{SU(6) \mbox{ rep}} & & \underline{SU(3) \mbox{ content reps}}
\nonumber \\
{\bf 35}   & & {\bf 1}, {\bf 8} \nonumber \\
{\bf 405}  & & {\bf 8}, {\bf 8}, {\bf 10}, \overline{\bf 10}, {\bf 27}
\nonumber \\
{\bf 2695} & & {\bf 1}, {\bf 8}, {\bf 8}, {\bf 10}, \overline{\bf 10},
{\bf 27}, {\bf 27}, {\bf 27}, {\bf 35}, \overline{\bf 35}, {\bf 64}
\hspace{2em}, \\ & & \nonumber \\
& \underline{J=2} & \nonumber \\
\underline{SU(6) \mbox{ rep}} & & \underline{SU(3) \mbox{ content reps}}
\nonumber \\
{\bf 405}  & & {\bf 1}, {\bf 8}, {\bf 27} \nonumber \\
{\bf 2695} & & {\bf 8}, {\bf 8}, {\bf 10}, \overline{\bf 10}, {\bf
27}, {\bf 27}, {\bf 35}, \overline{\bf 35}, {\bf 64} \hspace{2em},
\\ & & \nonumber \\
& \underline{J=3} & \nonumber \\
\underline{SU(6) \mbox{ reps}} & & \underline{SU(3) \mbox{ content reps}}
\nonumber \\
{\bf 2695} & & {\bf 1}, {\bf 8}, {\bf 27}, {\bf 64} \hspace{2em}.
\end{eqnarray}
Using that the $SU(3)$ reps ${\bf 1}$, ${\bf 8}$, ${\bf 10}$,
$\overline{\bf 10}$, ${\bf 27}$, ${\bf 35}$, $\overline{\bf 35}$, and
${\bf 64}$ respectively have 1, 2, 1, 1, 3, 2, 2, and 4 states with
$\Delta I_3 = \Delta Y = 0$, we count 92 chiral coefficients in total,
as expected, and numbers for each value of $J$ that agree with the
block-diagonalization counting for baryon bilinears given above.

	In order to implement tensor methods, we must have tensor
forms for both the ${\bf 35}$ and ${\bf 56}$.  As previously stated,
the ${\bf 35}$ may be represented as a traceless $6 \times 6$ matrix;
however, the trace adds only a harmless singlet to our analysis, and
so to obtain arbitrary second-order $SU(6)$ breaking, we require two
arbitrary $SU(6)$ matrices $X$ and $Z$.  The quantity we must compute
is $\overline{B}BXZ$, where $B$ is the tensor form of the ${\bf 56}$,
and $SU(6)$ indices are contracted in al possible ways.  In fact, the
very useful tensor $B$ appears in the literature~\cite{Pais}:

	We first define the familiar $SU(3)$ tensors.  For the baryon
octet,
\begin{equation}
O_a{}^b \equiv \left( \begin{array}{ccc} \label{m1}
\frac{1}{\sqrt{2}} \Sigma^0 + \frac{1}{\sqrt{6}} \Lambda & \Sigma^+ & p \\
\Sigma^- & -\frac{1}{\sqrt{2}} \Sigma^0 + \frac{1}{\sqrt{6}} \Lambda & n \\
\Xi^- & \Xi^0 & -\frac{2}{\sqrt{6}} \Lambda
\end{array} \right) .
\end{equation}
The baryon decuplet in this notation, a $3 \times 3 \times 3$ array, may be
represented as a collection of three matrices:
\begin{eqnarray} \label{m3}
T^{abc} & \equiv &
    \left(
    \hspace{-0.5em}
    \begin{array}{ccc}
\Delta^{++} & \frac{1}{\sqrt{3}} \Delta^+ & \frac{1}{\sqrt{3}} \Sigma^{*+} \\
\frac{1}{\sqrt{3}} \Delta^+ & \frac{1}{\sqrt{3}} \Delta^0 &
\frac{1}{\sqrt{6}} \Sigma^{*0} \\
\frac{1}{\sqrt{3}} \Sigma^{*+} & \frac{1}{\sqrt{6}} \Sigma^{*0} &
\frac{1}{\sqrt{3}} \Xi^{*0}
    \end{array}
    \hspace{-0.5em}
    \right)
    \hspace{-0.5em}
    \left(
    \hspace{-0.5em}
    \begin{array}{ccc}
\frac{1}{\sqrt{3}} \Delta^+ & \frac{1}{\sqrt{3}} \Delta^0 &
\frac{1}{\sqrt{6}} \Sigma^{*0} \\
\frac{1}{\sqrt{3}} \Delta^0 & \Delta^- & \frac{1}{\sqrt{3}} \Sigma^{*-} \\
\frac{1}{\sqrt{6}} \Sigma^{*0} & \frac{1}{\sqrt{3}} \Sigma^{*-} &
\frac{1}{\sqrt{3}} \Xi^{*-}
    \end{array}
    \hspace{-0.5em}
    \right)
    \hspace{-0.5em}
    \left(
    \hspace{-0.5em}
    \begin{array}{ccc}
\frac{1}{\sqrt{3}} \Sigma^{*+} & \frac{1}{\sqrt{6}} \Sigma^{*0} &
\frac{1}{\sqrt{3}} \Xi^{*0} \\
\frac{1}{\sqrt{6}} \Sigma^{*0} & \frac{1}{\sqrt{3}} \Sigma^{*-} &
\frac{1}{\sqrt{3}} \Xi^{*-} \\
\frac{1}{\sqrt{3}} \Xi^{*0} & \frac{1}{\sqrt{3}} \Xi^{*-} & \Omega^-
    \end{array}
    \hspace{-0.5em}
    \right) .
\end{eqnarray}
One may assign any particular permutation of indices {\it a,b,c\/} to denote
row, column, and sub-matrix in this representation, because the decuplet is
completely symmetric under rearrangement of flavor indices.

	Using the notation $\Uparrow$, $\uparrow$, $\downarrow$,
$\Downarrow$ to denote $J_3 = +\f{3}{2}, +\f{1}{2}, -\f{1}{2},
-\f{3}{2}$, the $SU(2)$ spin tensors for spin-1/2 and spin-3/2 assume
the forms
\begin{equation}
\chi^i \equiv \left( \begin{array}{c} \uparrow \\ \downarrow
\end{array} \right)
\end{equation}
and
\begin{equation}
\chi^{ijk} \equiv \begin{array}{cc}
\left( \begin{array}{cc} \Uparrow & \f{1}{\s{3}} \uparrow \\
\f{1}{\s{3}} \uparrow & \f{1}{\s{3}} \downarrow \end{array} \right) &
\left( \begin{array}{cc} \f{1}{\s{3}} \uparrow & \f{1}{\s{3}} \downarrow \\
\f{1}{\s{3}} \downarrow & \Downarrow \end{array} \right) \end{array} ,
\end{equation}
where the latter tensor is symmetric under exchange of indices.

	Then, with the use of the Levi-Civita symbols $\epsilon^{ij}$
and $\epsilon^{ijk}$, we construct the ${\bf 56}$ tensor:
\begin{eqnarray}
B^{aibjck} & = & \chi^{ijk} T^{abc} + \f{1}{3\s{2}} \left[
\epsilon^{ij} \chi^k \epsilon^{abd} O_d{}^c +
\epsilon^{jk} \chi^i \epsilon^{bcd} O_d{}^a +
\epsilon^{ki} \chi^j \epsilon^{cad} O_d{}^b \right] .
\end{eqnarray}
Note that $B$ is completely symmetric under the exchange of pairs of
indices from $SU(3) \times SU(2)$, as the ${\bf 56}$ is a symmetric
rep of $SU(6)$.  The $1/3\s{2}$ guarantees the singlet normalization:
\begin{equation}
\overline{B}_{aibjck} B^{aibjck} = \overline{p \! \uparrow} p \! \uparrow +
\overline{p \! \downarrow} p \! \downarrow +
\overline{\Delta^{+ \! +} \! \Uparrow} \Delta^{+ \! +} \! \Uparrow +
\overline{\Delta^{+ \! +} \! \uparrow} \Delta^{+ \! +} \! \uparrow + \cdots
\end{equation}

	Because we are interested in bilinear combinations with
definite $J$, we also require a table of $SU(2)$ Clebsch--Gordan
coefficients; however, since we have abandoned the Condon--Shortley
phase convention for the $SU(3)$ coefficients, we must do likewise for
their $SU(2)$ analogues.  Starting with Clebsches in the
Condon--Shortley convention, we choose all Clebsches $\langle 0 \, 0 |
s \, +m; \; s \, -m \rangle$ to be the same regardless of $m$, and both
values of $\langle 1 \, 0 | \f{3}{2} \, +m; \; \f{1}{2} \, -m
\rangle$ to be positive.  The $SU(2)$ relation
\begin{equation}
\langle j_1 \, \mbox{} +m_1 ; \; j_2 \, \mbox{} +m_2 | j \, \mbox{}
+m \rangle =
\langle j_2 \, \mbox{} -m_2 ; \; j_1 \, \mbox{} -m_1 | j \, \mbox{}
-m \rangle
\end{equation}
relates the $\f{3}{2} \times \f{1}{2}$ and $\f{1}{2} \times \f{3}{2}$
Clebsch tables.

	To obtain the $SU(6)$ Clebsch--Gordan coefficients in the
${\bf 35}$ rep, we simply compute the quantity $\overline{B} BXZ$ with
$X$ traceless and $Z={\bf 1}$.  To decompose into the component $SU(3)
\times SU(2)$ quantum numbers, we choose $X$ to consist of the basis
operators ${\bf 1} \otimes \sigma_3$, $Y \otimes {\bf 1}$, $I_3
\otimes {\bf 1}$, $Y \otimes \sigma_3$, and $I_3 \otimes \sigma_3$.
The $SU(6)$ rep ${\bf 1}$ is even more trivial: $X=Z={\bf 1}$.

	One may use a similar approach for ${\bf 405}$ and ${\bf
2695}$ operators as well, but then one must render the products of $6
\times 6$ matrices completely traceless under any contraction, and
this procedure tends to be tedious for larger reps in $SU(3) \times
SU(2)$ notation.  A much better approach is to find the ${\bf 2695}$
combinations by observing that it is exactly these combinations that
vanish in the quantity $\overline{B}BXZ$.  We know from the $SU(3)
\times SU(2)$ contents which reps appear, and we know from Sec.\ 2
that a particular $SU(6)$ chiral coefficient $d^{R,I}_N$ is simply a
linear combination of $SU(3)$ chiral coefficients with the same
quantum numbers $R,I$.  Therefore, we form an arbitrary linear
combination of the desired $SU(3)$ chiral coefficients and seek out
values of the coefficients for which this combination vanishes from
$\overline{B}BXZ$; such a combination transforms under the ${\bf
2695}$ rep.  If there are more than one, we arbitrarily choose an
orthogonalization to lift the degeneracy.  Finally, we find the chiral
coefficients $d^{R,I}_{405}$ by their orthogonality to
$d^{R,I}_{2695}$, $d^{R,I}_{35}$, and $d^{R,I}_1$.

	This procedure gives us all of the $SU(6)$ Clebsch--Gordan
coefficients for product states in $\overline{\bf 56} \otimes {\bf
56}$ with $\Delta I_3 = \Delta Y = \Delta J_3 = 0$.  As we have
pointed out, the restriction $\Delta J_3 = 0$ is of no great
consequence, for we may use the Wigner--Eckart theorem to obtain
coefficients with $\Delta J_3 \neq 0$.  $\Delta I_3, \Delta Y \neq 0$
are not much harder; because $SU(3)$ Clebsch--Gordan coefficients are
also well-known, we may use the $SU(3)$ version of the Wigner--Eckart
theorem to obtain the others.  Thus {\em all\/} coefficients of this
product are now known.  The great advantage of this approach is that
similar techniques may be applied to other product reps and other
symmetry groups.
\section{Exhibition of $SU(6)$ Clebsch--Gordan Coefficients}
	Here we collect the mathematical results of the procedure just
described in a compact notation.  Rather than exhibiting the gigantic
$92 \times 92$ matrix or even the smaller diagonal blocks, we present
sub-blocks associated with each $SU(3)$ rep ${\bf R}$.  Note
especially that the chiral coefficients $d^{R,I}_N$, for given $R$ and
$N$, are independent of the particular value of $I$.  On the other
hand, these coefficients depend implicitly upon $J$; when confusion
could arise, we write $d^{R,I}_{N,J}$.
\begin{eqnarray}
\underline{J=0} & & \nonumber \\ & &
\begin{array}{ccccccccc}
\left( \begin{array}{c} d^{1,0}_1 \\ d^{1,0}_{405} \end{array} \right) &
= & \left( \begin{array}{cc} +\s{\f{2}{7}} & +\s{\f{5}{7}} \\
+\s{\f{5}{7}} & -\s{\f{2}{7}} \end{array} \right) & \hspace{-0.3em}
\left( \begin{array}{c} a^1_0 \\ b^1_0 \end{array} \right) &
\hspace{2em} & \left( \begin{array}{c} d^{27,I}_{405} \\
d^{27,I}_{2695} \end{array} \right) & = &
\left( \begin{array}{cc} +\f{1}{\s{15}} & +\s{\f{14}{15}} \\
+\s{\f{14}{15}} & -\f{1}{\s{15}} \end{array} \right) & \hspace{-0.3em}
\left( \begin{array}{c} a^{27}_I \\ b^{27}_I \end{array} \right)
\end{array} \nonumber \\ & & \nonumber \\ & &
\begin{array}{cccc} \left( \begin{array}{c} d^{8,I}_{35} \\
d^{8,I}_{405} \\ d^{8,I}_{2695} \end{array} \right) & = & \left(
\begin{array}{ccc} 0 & +\f{1}{\s{6}} & +\s{\f{5}{6}} \\ +\s{\f{2}{5}}
& +\f{1}{\s{2}} & -\f{1}{\s{10}} \\ +\s{\f{3}{5}} & -\f{1}{\s{3}} &
+\f{1}{\s{15}} \end{array} \right) & \hspace{-0.3em} \left(
\begin{array}{c} a^{8_1}_I
\\ a^{8_2}_I \\ b^8_I \end{array} \right) \end{array} \nonumber \\ & &
\nonumber \\ & &
d^{10,1}_{2695} = a^{10}_1 \hspace{2em} d^{\overline{10},1}_{2695} =
a^{\overline{10}}_1 \hspace{2em} d^{64,I}_{2695} = b^{64}_I,
\end{eqnarray}
\begin{eqnarray}
\underline{J=1} & & \nonumber \\ & &
\begin{array}{cccc}
\left( \begin{array}{c} d^{1,0}_{35} \\ d^{1,0}_{2695} \end{array}
\right) & = & \left( \begin{array}{cc} +\f{\s{2}}{3\s{3}} &
+\f{5}{3\s{3}} \\ +\f{5}{3\s{3}} & -\f{\s{2}}{3\s{3}} \end{array}
\right) & \hspace{-0.3em} \left( \begin{array}{c} a^1_0 \\ b^1_0
\end{array} \right) \end{array} \nonumber \\ & & \nonumber \\ & &
\begin{array}{cccc}
\left( \begin{array}{c} d^{8,I}_{35} \\ d^{8_1,I}_{405} \\
d^{8_2,I}_{405} \\ d^{8_1,I}_{2695} \\ d^{8_2,I}_{2695} \end{array}
\right) & = & \left( \begin{array}{ccccc} -\f{\s{5}}{3\s{6}} &
+\f{\s{2}}{3\s{3}} & +\f{5}{3\s{6}} & +\f{\s{5}}{3\s{3}} &
+\f{\s{5}}{3\s{3}} \\ +\f{1}{\s{10}} & 0 & +\f{1}{\s{2}} &
-\f{1}{\s{5}} & -\f{1}{\s{5}} \\ 0 & 0 & 0 & +\f{1}{\s{2}} &
-\f{1}{\s{2}} \\ +\f{2}{\s{5}} & 0 & 0 & +\f{1}{\s{10}} &
+\f{1}{\s{10}} \\ +\f{1}{3\s{15}} & +\f{5}{3\s{3}} & -\f{1}{3\s{3}} &
-\f{\s{2}}{3\s{15}} & -\f{\s{2}}{3\s{15}} \end{array} \right) &
\hspace{-0.3em} \left(
\begin{array}{c} a^{8_1}_I \\ a^{8_2}_I \\ b^8_I \\ c^8_I \\
\overline{c}^8_I \end{array} \right) \end{array} \nonumber \\
\vspace{1ex} & & \nonumber \\ & &
\begin{array}{ccccccccc} \left( \begin{array}{c} d^{10,1}_{405} \\
d^{10,1}_{2695} \end{array} \right) & = & \left( \begin{array}{cc}
+\f{1}{\s{5}} & +\f{2}{\s{5}} \\ +\f{2}{\s{5}} & -\f{1}{\s{5}}
\end{array} \right) & \hspace{-0.3em} \left( \begin{array}{c} a^{10}_1
\\ c^{10}_1 \end{array} \right) & \hspace{2em} & \left(
\begin{array}{c} d^{\overline{10},1}_{405} \\
d^{\overline{10},1}_{2695} \end{array} \right) & = & \left(
\begin{array}{cc} +\f{1}{\s{5}} & +\f{2}{\s{5}} \\ +\f{2}{\s{5}} &
-\f{1}{\s{5}} \end{array} \right) & \hspace{-0.3em} \left(
\begin{array}{c} a^{\overline{10}}_1 \\ c^{\overline{10}}_1
\end{array} \right) \end{array} \nonumber \\ & & \nonumber \\ & &
\begin{array}{cccc}
\left( \begin{array}{c} d^{27,I}_{405} \\ d^{27_1,I}_{2695} \\
d^{27_2,I}_{2695} \\ d^{27_3,I}_{2695} \end{array} \right) & = &
\left( \begin{array}{cccc} +\f{\s{2}}{3\s{5}} & +\f{\s{7}}{3} &
+\f{2}{3\s{5}} & +\f{2}{3\s{5}} \\ +\f{2}{\s{5}} & 0 & -\f{1}{\s{10}}
& -\f{1}{\s{10}} \\ +\f{\s{7}}{3\s{5}} & -\f{\s{2}}{3} &
+\f{\s{14}}{3\s{5}} & +\f{\s{14}}{3\s{5}} \\ 0 & 0 & +\f{1}{\s{2}} &
-\f{1}{\s{2}} \end{array} \right) & \hspace{-0.3em} \left(
\begin{array}{c} a^{27}_I \\ b^{27}_I \\ c^{27}_I \\ \overline{c}^{27}_I
\end{array} \right) \end{array} \nonumber \\ & & \nonumber \\ & &
d^{35,I}_{2695} = c^{35}_I \hspace{2em} d^{\overline{35},I}_{2695} =
\overline{c}^{\overline{35}}_I \hspace{2em} d^{64,I}_{2695} = b^{64}_I,
\end{eqnarray}
\begin{eqnarray}
\underline{J=2} & & \nonumber \\ & &
d^{1,0}_{405} = b^1_0 \nonumber \\ & & \nonumber \\ & &
\begin{array}{cccc} \left( \begin{array}{c} d^{8,I}_{405} \\
d^{8_1,I}_{2695} \\ d^{8_2,1}_{2695} \end{array} \right) & = & \left(
\begin{array}{ccc} +\f{2}{\s{5}} & +\f{1}{\s{10}} & +\f{1}{\s{10}} \\
+\f{1}{\s{5}} & -\s{\f{2}{5}} & -\s{\f{2}{5}} \\ 0 & +\f{1}{\s{2}} &
-\f{1}{\s{2}} \end{array} \right) & \hspace{-0.3em} \left(
\begin{array}{c} b^8_I \\ c^8_I \\ \overline{c}^8_I \end{array}
\right) \end{array} \nonumber
\\ & & \nonumber \\ & & d^{10,1}_{2695} = c^{10}_1 \hspace{2em}
d^{\overline{10},1}_{2695} = \overline{c}^{\overline{10}}_1 \nonumber
\\ & & \nonumber \\ & &
\begin{array}{cccc} \left( \begin{array}{c} d^{27,I}_{405} \\
d^{27_1,I}_{2695} \\ d^{27_2,I}_{2695} \end{array} \right) & = &
\left( \begin{array}{ccc} +\s{\f{7}{15}} & +\f{2}{\s{15}} &
+\f{2}{\s{15}} \\ +\s{\f{8}{15}} & -\s{\f{7}{30}} & -\s{\f{7}{30}} \\
0 & +\f{1}{\s{2}} & -\f{1}{\s{2}} \end{array} \right) &
\hspace{-0.3em} \left( \begin{array}{c} b^{27}_I \\ c^{27}_I \\
\overline{c}^{27}_I \end{array} \right) \end{array} \hspace{10em}
\mbox{ } \nonumber \\
& & \nonumber \\ & &
d^{35,I}_{2695} = c^{35}_I \hspace{2em} d^{\overline{35},I}_{2695} =
\overline{c}^{\overline{35}}_I \hspace{2em} d^{64,I}_{2695} = b^{64}_I,
\end{eqnarray}
\hspace{3em} $\underline{J=3}$
\begin{equation}
d^{R,I}_{2695} = b^R_I \hspace{2em} \mbox{for $R$ = 1, 8, 27, 64.}
\end{equation}

	A number of chiral coefficients contain redundant information
because of symmetry under conjugation ({\it e.g.\/} $d^{10,1}_{N,J}$ and
$d^{\overline{10},1}_{N,J}$).  Others ({\it e.g.\/} $d^{8_2,I}_{405,1}$)
necessarily vanish once we impose hermiticity.  These are the
counterparts to the degrees of freedom lost from demanding only
Hermitian combinations of bilinears like ($\overline{\Lambda}
\Sigma^{*0} + \overline{\Sigma^{*0}} \Lambda$), and one
finds, as expected, exactly 74 physically independent chiral
coefficients.

	In order to obtain baryon relations, we must take into account
the particular matrix elements used in defining the mass and
electromagnetic moments.  The matrices above are defined by bilinears
in eigenstates of total $J$, but the various moments are defined as
matrix elements connecting the states with {\em highest weight\/} in
the spin-projection quantum number.  The magnetic dipole moment of a
particle with spin $s$, for example, is defined as the matrix element
with angular momentum structure $\langle 1 \, 0 | s \, -s; \; s \,
\mbox{} +s \rangle$.  In the case of transitions between particles
with different spins $s_1$ and $s_2$, however, the convention is not
so universal.  We adopt the choice that the two particles are taken to
be in the highest-weight spin states such that their combined value is
still zero; that is, the spin-$J$ multipole moment transition is
defined through the matrix element
\begin{equation}
\langle J \, 0 | s_1 \, \mbox{} -\min (s_1,s_2) ; \; s_2 \, \mbox{}
+\min (s_1,s_2) \rangle .
\end{equation}
Note that the $J=0$ matrix elements, which give rise to masses (or
electric charges), do not depend on this choice because of our
previous choice of Clebsch--Gordan convention; here the physical fact
of the independence of baryon masses on individual spin states becomes
most clear.  The matrix elements for all multipole moments can now be
obtained trivially from the $SU(6)$ matrices by use of the $SU(2)$
Wigner--Eckart theorem.
\section{Baryon Relations}
\subsection{Estimating Relation-breaking Terms}
	As in Sec.\ 3 for the case of $SU(3)$, we argue that the
largest reps of $SU(6)$ give rise to the most experimentally accurate
relations.  $SU(6)$-breaking operators appear in the small ${\bf 35}$
rep; the largest rep, the ${\bf 2695}$, requires three of these in
product, so the ${\bf 2695}$ chiral coefficients contain all relations
third order in $SU(6)$ breaking.  The only statement that must be
verified is that all of the ${\bf 35}$ operators possess numerically
small coefficients.  Certainly the quark mass and charge operators,
now written in $SU(3) \times SU(2)$ notation as $M_q \otimes {\bf 1}$
and $Q_q \otimes {\bf 1}$, are still small, as are the corresponding
operators with with spin-flips, $M_q \otimes \sigma_3$ and $Q_q
\otimes \sigma_3$.  The only other physical operator to consider is
the pure spin-flip ${\bf 1} \otimes \sigma_3$.  {\it A priori\/} we
see no reason this operator should have a small coefficient, but it is
precisely this operator that explains the relative smallness of the
breaking between the average octet and decuplet baryon masses.  Thus
even this operator must possess a numerically small coefficient.

	In order to judge the quality of the following relations, we
must be able to estimate the coefficients of these ${\bf 2695}$
operators.  Fortunately, this is a matter of simple naive dimensional
analysis; we assume that any unknown dimensionless parameters are of
order one.  For simplicity, let us consider the mass relations only.
The numerical breaking of average octet and decuplet masses can be
characterized by the number
\begin{equation}
\f{m_{10} - m_8}{\f{1}{2}(m_{10} + m_8)} \approx 0.2 .
\end{equation}
We use this to estimate the spin-flip coefficient conservatively as
$0.3$.  Therefore, $I=0$ operators in the ${\bf 2695}$ contribute an
amount to each baryon mass of order
\begin{equation}
\Lambda_{\chi} (0.3)^3 \approx 25 \mbox{ MeV}.
\end{equation}
Isospin-breaking operators are much more heavily suppressed.  Each
unit of isospin breaking contributes an additional factor $(m_d -
m_u)/\Lambda_{\chi} \approx 0.005$; alternately, for each {\em two\/}
units of isospin breaking, a factor of $\alpha/4\pi$ can appear
(Operators with single powers of $e$ are forbidden in masses by charge
conjugation symmetry).  Typical numbers are 0.5, 0.2, and 0.003 MeV
for $I=1,2$, and 3, respectively.  Note that these naive estimates
apply to {\em individual\/} baryons, and large coefficients in the
relations presented below must be taken into account to obtain
reliable numbers.  Similar arguments apply to the electromagnetic
moment relations.
\subsection{Masses}
	Here we exhibit the mass combinations associated with each
chiral coefficient in the ${\bf 2695}$.  There are 19 independent
parameters in the $J=0$ sector, corresponding to the octet and decuplet
masses and the $\Sigma^0$-$\Lambda$ mixing parameter, which we denote
by $\beta$.  The ten $J=0$ chiral coefficients in the ${\bf 2695}$,
characterized by their $SU(3)$ decompositions, are
\begin{eqnarray}
\underline{(SU(3),I)} & \hspace{1em} & \underline{\mbox{Mass
combination}} \nonumber \\
(8,0) & & \mbox{} +(p+n) + 3(\Sigma^+ + \Sigma^0 + \Sigma^-)
- 3\Lambda - 4(\Xi^0 + \Xi^-) \nonumber \\
& & -(\Delta^{+ \! +} + \Delta^+ + \Delta^0 + \Delta^-) +
(\Xi^{*0} + \Xi^{*-}) + 2\Omega^- , \nonumber \\
(8,1) & & \mbox{} +7(p-n) + 5(\Sigma^+ - \Sigma^-) - 2(\Xi^0 - \Xi^-)
-6\s{3} \beta \nonumber \\
& & -(3\Delta^{+ \! +} + \Delta^+ - \Delta^0 - 3\Delta^-) -
2(\Sigma^{*+} - \Sigma^{*-}) -(\Xi^{*0} - \Xi^{*-}) , \nonumber \\
(27,0) & & \mbox{} +7[3(p+n) - (\Sigma^+ + \Sigma^0 + \Sigma^-) -9\Lambda +
3(\Xi^0 + \Xi^-)] \nonumber \\
& & - (\Delta^{+ \! +} + \Delta^+ + \Delta^0 + \Delta^-) +
(\Xi^{*0} + \Xi^{*-}) + 2\Omega^- , \nonumber \\
(27,1) & & \mbox{} +7[(p-n) -(\Xi^0 - \Xi^-) + 2\s{3} \beta ]
\nonumber \\ & & - (3\Delta^{+ \! +} + \Delta^+ - \Delta^0 - 3\Delta^-)
+ 3(\Sigma^{*+} - \Sigma^{*-}) + 4(\Xi^{*0} - \Xi^{*-}) , \nonumber \\
(27,2) & & \mbox{} +7(\Sigma^+ -2\Sigma^0 + \Sigma^-) \nonumber \\
& & -3(\Delta^{+ \! +} - \Delta^+ - \Delta^0 + \Delta^-) -(\Sigma^{*+}
- 2\Sigma^{*0} + \Sigma^{*-}) , \nonumber \\
(10,1),(\overline{10},1) & & \mbox{} +(p-n) -(\Sigma^+ - \Sigma^-) +
(\Xi^0 - \Xi^-) , \nonumber \\
(64,0) & & \mbox{} + (\Delta^{+ \! +} + \Delta^+ + \Delta^0 + \Delta^-) -
4(\Sigma^{*+} + \Sigma^{*0} + \Sigma^{*-}) + 6(\Xi^{*0} + \Xi^{*-}) -
4\Omega^- , \nonumber \\
(64,1) & & \mbox{} + (3\Delta^{+ \! +} + \Delta^+ - \Delta^0 -
3\Delta^-) - 10(\Sigma^{*+} - \Sigma^{*-}) + 10(\Xi^{*0} - \Xi^{*-}) ,
\nonumber \\
(64,2) & & \mbox{} +(\Delta^{+ \! +} - \Delta^+ - \Delta^0 + \Delta^-) -
2(\Sigma^{*+} - 2\Sigma^{*0} + \Sigma^{*-}) , \nonumber \\
(64,3) & & \mbox{} +\Delta^{+ \! +} - 3\Delta^+ + 3\Delta^0 - \Delta^- .
\end{eqnarray}
It is interesting to note that the last five of these are also $SU(3)$
relations as well, because the $SU(3)$ reps ${\bf 10}$ and
$\overline{\bf 10}$ do not appear in the decuplet-decuplet product,
and ${\bf 64}$ does not appear in the octet-octet product.  In fact,
since the ${\bf 64}$ rep neither appears in ${\bf 8} \otimes {\bf 10}$
nor $\overline{\bf 10} \otimes {\bf 8}$, we have the curious result
that the these combinations of decuplet bilinears give not only mass
but dipole, quadrupole, and octupole moment relations with the same
coefficients.

	We also point out that the three $I=0$ relations, for which we
may neglect mass differences within each isospin multiplet, are
equivalent to the three relations derived by Dashen, Jenkins, and
Manohar~\cite{DJM} in the large-$N_c$ contracted spin-flavor symmetry.
This is an excellent illustration of the similarity between the two
symmetries.

	We now exhibit numerical values for these combinations.  In
all cases we use Particle Data Group (PDG)~\cite{PDG} values for the
masses, with the following exceptions.  First, the unknown parameter
$\beta$ is eliminated between the $(8,1)$ and $(27,1)$ combinations.
Next, the $\Delta$ mass differences have notoriously large
uncertainties; we adopt the arguments in Ref.~\cite{L2} that a set
consistent with chiral loop calculations is
\begin{eqnarray}
\Delta^0 - \Delta^{+ \! +} & = & 1.3 \pm 0.5 \mbox{ MeV}, \nonumber \\
\Delta^+ & = & 1231.5 \pm 0.3 \mbox{ MeV}.
\end{eqnarray}
{}From the same reference, we fix the $\Delta^-$ mass, which has never
been directly determined, by means of the $(64,3)$ relation; its
corrections, including loop effects, are determined to be negligible.
The results are presented in Table I\@.  In all cases, the naive
estimates of ${\bf 2695}$ operators explain the experimental relation
breakings.

	The set of nine relations after the elimination of $\beta$ is
equivalent to the set derived by Rubinstein, Scheck, and
Socolow~\cite{RSS}, who used very similar reasoning to that above;
their neglect of ``three-body operators'' is equivalent to the neglect
of the ${\bf 2695}$.  The difference is that the earlier authors did
not distinguish the relations by $SU(3)$ content.  On one hand,
their $SU(3)$ decomposition of the ${\bf 2695}$ is missing the
$(10,1)$ and $(\overline{10},1)$ terms (one independent parameter),
and on the other the $\Sigma^0$-$\Lambda$ mixing is neglected; thus
they count only nine relations.

	This brings us to the tenth relation, that which predicts
$\beta$.  We choose the unique sum of $(8,1)$ and $(27,1)$ that
eliminates the troublesome $\Delta$ masses, and obtain the pretty
result
\begin{eqnarray}
\beta & = & \mbox{} +\f{1}{4\s{3}} [(\Sigma^+ - \Sigma^-) + (\Xi^0 -
\Xi^-) - (\Sigma^{*+} - \Sigma^{*-}) - (\Xi^{*0} - \Xi^{*-})] \nonumber
\\ & = & -0.99 \pm 0.15 \mbox{ MeV}.
\end{eqnarray}
A naive estimate of the ${\bf 2695}$ breaking of this relation
produces a further uncertainty of order 0.2--0.3 MeV.  It is important
to recognize that the masses above labeled $\Sigma^0$ and $\Lambda$
are actually eigenvalues associated with isospin eigenstates; to
obtain the mass eigenvalues, we must diagonalize a $2 \times 2$ matrix
including the mixing terms~\cite{FLNC}.  If we define the mass
eigenvalues (labeled by $m$) via
\begin{equation}
\left( \begin{array}{c} \Sigma^0_m \\ \Lambda_m \end{array} \right) =
\left( \begin{array}{cc} +\cos \theta & +\sin \theta \\ -\sin \theta &
+\cos \theta \end{array} \right) \left( \begin{array}{c} \Sigma^0 \\
\Lambda \end{array} \right) ,
\end{equation}
then we find
\begin{equation}
\theta = -0.013 \pm 0.002 \mbox{ rad} \hspace{2em} (-0.74 \pm 0.11^\circ ),
\end{equation}
where
\begin{equation}
\theta = \f{1}{2} \tan^{-1} \left[ \f{\beta}{\s{\left(\f{\Sigma^0_m -
\Lambda_m}{2} \right)^2 + \beta^2}} \right] .
\end{equation}
The difference $(\Sigma^0_m - \Sigma^0) = -(\Lambda_m - \Lambda)$
turns out to be a mere $13 \pm 4$ keV, and thus we lose nothing by
using mass eigenvalues for $\Sigma^0$ and $\Lambda$ in the other mass
relations.
\subsection{Magnetic Dipole Moments}
	The $J=1$ sector is characterized by 27 parameters, which may
be thought of as the magnetic dipole moments of the octet and decuplet
baryons, the eight possible transition moments between these
multiplets, and the $\Sigma^0$-$\Lambda$ transition moment.  There are
18 independent chiral coefficients in the ${\bf 2695}$, given by
\begin{eqnarray}
\underline{(SU(3),I)} & \hspace{1em} & \underline{\mbox{Magnetic
dipole moment combination}} \nonumber \\
(1,0) & & \mbox{} +15[(\mu_p + \mu_n) + (\mu_{\Sigma^+} +
\mu_{\Sigma^0} + \mu_{\Sigma^-}) +\mu_{\Lambda} + (\mu_{\Xi^0} +
\mu_{\Xi^-})] \nonumber \\ & &
\mbox{} -4[(\mu_{\Delta^{+ \! +}} + \mu_{\Delta^+} +
\mu_{\Delta^0} + \mu_{\Delta^-}) + (\mu_{\Sigma^{*+}} +
\mu_{\Sigma^{*0}} + \mu_{\Sigma^{*-}}) \nonumber \\ & &
\mbox{} \hspace{2em} \mbox{} + (\mu_{\Xi^{*0}} + \mu_{\Xi^{*-}}) +
\mu_{\Omega^-}], \nonumber \\
(8_1,0) & & \mbox{} +[(\mu_p + \mu_n) - 2(\mu_{\Sigma^+} +
\mu_{\Sigma^0} + \mu_{\Sigma^-}) + 2\mu_{\Lambda} + (\mu_{\Xi^0} +
\mu_{\Xi^-})] \nonumber \\ & &
\mbox{} +\s{2} [(\mu_{\Sigma^+ \Sigma^{*+}} + \mu_{\Sigma^0
\Sigma^{*0}} + \mu_{\Sigma^- \Sigma^{*-}}) + (\mu_{\Xi^0 \Xi^{*0}} +
\mu_{\Xi^- \Xi^{*-}})], \nonumber \\
(8_1,1) & & \mbox{} +3[(\mu_p - \mu_n) - (\mu_{\Xi^0} - \mu_{\Xi^-})]
-4\s{3} \mu_{\Sigma^0 \Lambda} \nonumber \\ & &
\mbox{} -\s{2} [2(\mu_{p \Delta^+} + \mu_{n \Delta^0}) + (\mu_{\Sigma^+
\Sigma^{*+}} - \mu_{\Sigma^- \Sigma^{*-}}) + \s{3} \mu_{\Lambda
\Sigma^{*0}} \nonumber \\ & &
\mbox{} \hspace{2.5em} \mbox{}  + (\mu_{\Xi^0 \Xi^{*0}} -
\mu_{\Xi^- \Xi^{*-}})], \nonumber \\
(8_2,0) & & \mbox{} +3[13(\mu_p + \mu_n) - (\mu_{\Sigma^+} +
\mu_{\Sigma^0} + \mu_{\Sigma^-}) +\mu_{\Lambda} - 12(\mu_{\Xi^0} +
\mu_{\Xi^-})] \nonumber \\ & &
\mbox{} - 5[(\mu_{\Delta^{+ \! +}} + \mu_{\Delta^+} + \mu_{\Delta^0} +
\mu_{\Delta^-}) - (\mu_{\Xi^{*0}} + \mu_{\Xi^{*-}}) -
2\mu_{\Omega^-}] \nonumber \\ & &
\mbox{} -6\s{2} [(\mu_{\Sigma^+ \Sigma^{*+}} + \mu_{\Sigma^0
\Sigma^{*0}} + \mu_{\Sigma^- \Sigma^{*-}}) + (\mu_{\Xi^0 \Xi^{*0}} +
\mu_{\Xi^- \Xi^{*-}})] , \nonumber \\
(8_2,1) & & \mbox{} +3[11(\mu_p - \mu_n) + 25(\mu_{\Sigma^+} -
\mu_{\Sigma^-}) + 14(\mu_{\Xi^0} - \mu_{\Xi^-}) + 2\s{3} \mu_{\Sigma^0
\Lambda}] \nonumber \\ & &
\mbox{} -5[(3\mu_{\Delta^{+ \! +}} + \mu_{\Delta^+} - \mu_{\Delta^0} -
3\mu_{\Delta^-}) + 2(\mu_{\Sigma^{*+}} - \mu_{\Sigma^{*-}}) +
(\mu_{\Xi^{*0}} - \mu_{\Xi^{*-}})] \nonumber \\ & &
\mbox{} -6\s{2} [2(\mu_{p \Delta^+} + \mu_{n \Delta^0}) + (\mu_{\Sigma^+
\Sigma^{*+}} - \mu_{\Sigma^- \Sigma^{*-}}) + \s{3} \mu_{\Lambda
\Sigma^{*0}} \nonumber \\ & &
\mbox{} \hspace{3em} \mbox{} + (\mu_{\Xi^0 \Xi^{*0}} -
\mu_{\Xi^- \Xi^{*-}})], \nonumber \\
(10,1),(\overline{10},1) & & \mbox{} + 2[(\mu_p - \mu_n) -
(\mu_{\Sigma^+} - \mu_{\Sigma^-}) + (\mu_{\Xi^0} - \mu_{\Xi^-})]
\nonumber \\ & & \mbox{} -\s{2} [(\mu_{p \Delta^+} + \mu_{n \Delta^0})
- (\mu_{\Sigma^+ \Sigma^{*+}} - \mu_{\Sigma^- \Sigma^{*-}}) -
(\mu_{\Xi^0 \Xi^{*0}} - \mu_{\Xi^- \Xi^{*-}})], \nonumber \\
(27_1,0) & & \mbox{} +[3(\mu_p + \mu_n) - (\mu_{\Sigma^+} +
\mu_{\Sigma^0} + \mu_{\Sigma^-}) - 9\mu_{\Lambda} + 3(\mu_{\Xi^0} +
\mu_{\Xi^-})] \nonumber \\ & &
\mbox{} -\s{2} [2(\mu_{\Sigma^+ \Sigma^{*+}} + \mu_{\Sigma^0
\Sigma^{*0}} + \mu_{\Sigma^- \Sigma^{*-}}) - 3(\mu_{\Xi^0 \Xi^{*0}} +
\mu_{\Xi^- \Xi^{*-}})], \nonumber \\
(27_1,1) & & \mbox{} +2[(\mu_p - \mu_n) - (\mu_{\Xi^0} - \mu_{\Xi^-})
+ 2\s{3} \mu_{\Sigma^0 \Lambda}] \nonumber \\ & &
\mbox{} -5\s{2}[(\mu_{p \Delta^+} + \mu_{n \Delta^0}) + 3(\mu_{\Sigma^+
\Sigma^{*+}} - \mu_{\Sigma^- \Sigma^{*-}}) - 2\s{3} \mu_{\Lambda
\Sigma^{*0}} \nonumber \\ & &
\mbox{} \hspace{3.5em} \mbox{}  - 2(\mu_{\Xi^0 \Xi^{*0}} -
\mu_{\Xi^- \Xi^{*-}})], \nonumber \\
(27_1,2) & & \mbox{} + 4[\mu_{\Sigma^+} - 2\mu_{\Sigma^0} +
\mu_{\Sigma^-}] \nonumber \\ & &
\mbox{} -\s{2}[3(\mu_{p \Delta^+} - \mu_{n \Delta^0}) + (\mu_{\Sigma^+
\Sigma^{*+}} - 2\mu_{\Sigma^0 \Sigma^{*0}} + \mu_{\Sigma^-
\Sigma^{*-}})], \nonumber \\
(27_2,0) & & \mbox{} + 21[3(\mu_p + \mu_n) - (\mu_{\Sigma^+} +
\mu_{\Sigma^0} + \mu_{\Sigma^-}) - 9\mu_{\Lambda} + 3(\mu_{\Xi^0} +
\mu_{\Xi^-})] \nonumber \\ & &
\mbox{} - 20[3(\mu_{\Delta^{+ \! +}} + \mu_{\Delta^+} +
\mu_{\Delta^0} + \mu_{\Delta^-}) - 5(\mu_{\Sigma^{*+}} +
\mu_{\Sigma^{*0}} + \mu_{\Sigma^{*-}}) \nonumber \\ & &
\mbox{} \hspace{2em} \mbox{} - 3(\mu_{\Xi^{*0}} + \mu_{\Xi^{*-}}) +
9\mu_{\Omega^-}] \nonumber \\ & &
\mbox{} + 84\s{2} [2(\mu_{\Sigma^+ \Sigma^{*+}} + \mu_{\Sigma^0
\Sigma^{*0}} + \mu_{\Sigma^- \Sigma^{*-}}) - 3(\mu_{\Xi^0 \Xi^{*0}} +
\mu_{\Xi^- \Xi^{*-}})], \nonumber \\
(27_2,1) & & \mbox{} + 21[(\mu_p - \mu_n) - (\mu_{\Xi^0} - \mu_{\Xi^-})
+ 2\s{3} \mu_{\Sigma^0 \Lambda}] \nonumber \\ & &
\mbox{} - 10[(3\mu_{\Delta^{+ \! +}} + \mu_{\Delta^+} - \mu_{\Delta^0} -
3\mu_{\Delta^-}) - 3(\mu_{\Sigma^{*+}} - \mu_{\Sigma^{*-}}) -
4(\mu_{\Xi^{*0}} - \mu_{\Xi^{*-}})] \nonumber \\ & &
\mbox{} + 21\s{2}[(\mu_{p \Delta^+} + \mu_{n \Delta^0}) + 3(\mu_{\Sigma^+
\Sigma^{*+}} - \mu_{\Sigma^- \Sigma^{*-}}) - 2\s{3} \mu_{\Lambda
\Sigma^{*0}} \nonumber \\ & &
\mbox{} \hspace{3.5em} \mbox{}  - 2(\mu_{\Xi^0 \Xi^{*0}} -
\mu_{\Xi^- \Xi^{*-}})], \nonumber \\
(27_2,2) & & \mbox{} + 21[\mu_{\Sigma^+} - 2\mu_{\Sigma^0} +
\mu_{\Sigma^-}] \nonumber \\ & &
\mbox{} - 10[3(\mu_{\Delta^{+ \! +}} - \mu_{\Delta^+} - \mu_{\Delta^0}
+ \mu_{\Delta^-}) + (\mu_{\Sigma^{*+}} - 2\mu_{\Sigma^{*0}} +
\mu_{\Sigma^{*-}})] \nonumber \\ & &
\mbox{} + 21\s{2}[3(\mu_{p \Delta^+} - \mu_{n \Delta^0}) + (\mu_{\Sigma^+
\Sigma^{*+}} - 2\mu_{\Sigma^0 \Sigma^{*0}} + \mu_{\Sigma^-
\Sigma^{*-}})], \nonumber \\
(35,1),(\overline{35},1) & & \mbox{} +(\mu_{p \Delta^+} +
\mu_{n \Delta^0}) - (\mu_{\Sigma^+ \Sigma^{*+}} - \mu_{\Sigma^-
\Sigma^{*-}}) - 2\s{3} \mu_{\Lambda \Sigma^{*0}} \nonumber \\ & &
\mbox{} \hspace{1em} \mbox{} + 2(\mu_{\Xi^0 \Xi^{*0}} -
\mu_{\Xi^- \Xi^{*-}}), \nonumber \\
(35,2),(\overline{35},2) & & \mbox{} + (\mu_{p \Delta^+} - \mu_{n
\Delta^0}) - (\mu_{\Sigma^+ \Sigma^{*+}} - 2\mu_{\Sigma^0 \Sigma^{*0}}
+ \mu_{\Sigma^- \Sigma^{*-}}), \nonumber \\
(64,0) & & \mbox{} + (\mu_{\Delta^{+ \! +}} + \mu_{\Delta^+} +
\mu_{\Delta^0} + \mu_{\Delta^-}) - 4(\mu_{\Sigma^{*+}} +
\mu_{\Sigma^{*0}} + \mu_{\Sigma^{*-}}) \nonumber \\ & &
\mbox{} \hspace{1em} \mbox{} + 6(\mu_{\Xi^{*0}} + \mu_{\Xi^{*-}}) -
4\mu_{\Omega^-}, \nonumber \\
(64,1) & & \mbox{} + (3\mu_{\Delta^{+ \! +}} + \mu_{\Delta^+} -
\mu_{\Delta^0} - 3\mu_{\Delta^-}) - 10(\mu_{\Sigma^{*+}} -
\mu_{\Sigma^{*-}}) + 10(\mu_{\Xi^{*0}} - \mu_{\Xi^{*-}}),
\nonumber \\
(64,2) & & \mbox{} + (\mu_{\Delta^{+ \! +}} - \mu_{\Delta^+} -
\mu_{\Delta^0} + \mu_{\Delta^-}) - 2(\mu_{\Sigma^{*+}} -
2\mu_{\Sigma^{*0}} + \mu_{\Sigma^{*-}}), \nonumber \\
(64,3) & & \mbox{} + \mu_{\Delta^{+ \! +}} - 3\mu_{\Delta^+} +
3\mu_{\Delta^0} - \mu_{\Delta^-}.
\end{eqnarray}

	Ideally, because the decuplet and octet-decuplet dipole
moments are largely unknown, it would be preferable to have relations
written in terms of the octet only.  However, the only reps distinct
to a particular $SU(3)$ product in the $J=1$ sector are ${\bf 35}$ and
its conjugate (octet-decuplet transitions), and ${\bf 64}$ (decuplet
moments), and so such a reduction is impossible.  However, once we
assume the relations, there are only $27-18=9$ free moments, and
exactly this many are well-known; these are $\mu_{\Omega^-}$ and all
octet moments, including $\mu_{\Sigma^0 \Lambda}$, but not
$\mu_{\Sigma^0}$.  In terms of these, all 18 poorly-known or unknown
moments may be written.  The predictions are presented in Table II\@.

	Our prediction for the $\Delta^{+ \! +}$ dipole moment of $5.42
\pm 0.49 \mu_N$ is certainly consistent with the PDG estimate
$\mu_{\Delta{+ \! +}} = 3.7$ to $7.5 \mu_N$.  The only other known
dipole moment is $\mu_{p \Delta^+}$, which may be extracted from PDG
values for photon helicity amplitudes $A_{\f{1}{2},\f{3}{2}}$.  The
relation is
\begin{equation}
\mu_{p \Delta^+} = -\f{(A_{\f{1}{2}} + \s{3} A_{\f{3}{2}})}{\s{\pi
\alpha k}} \f{m_p}{1+\f{m_p}{m_{\Delta^+}}} \s{\f{m_p}{m_{\Delta^+}}}
\end{equation}
where $k$, the photon momentum in the decay, is fixed by kinematics.
This formula is obtained by comparing the amplitude for the decay in
terms of $\mu_{p \Delta^+}$ (see, {\it e.g.\/}, Ref.~\cite{RSS}) to
the same amplitude in terms of helicity amplitudes (see, {\it e.g.},
Ref.~\cite{hel}).  The PDG value is calculated to be $3.53 \pm 0.09
\mu_N$, in unfavorable comparison with our prediction of $2.52 \pm
0.23 \mu_N$.  The quark model, on the other hand, predicts $2.66
\mu_N$, whereas the large-$N_c$ contracted symmetry predicts~\cite{JM}
the much closer $3.33 \mu_N$ (Both of these predictions are functions
of $\mu_{p,n}$ only, and therefore have negligible uncertainties).
That the $SU(6)$ prediction is not closer to the experimental value
than the quark-model prediction is surprising, because $SU(6)$
contains the quark model, in a sense, as its lowest-order terms.  We
now describe this identification.

	Neglecting only the ${\bf 2695}$ means, of course, that the
fit to dipole moments is made using only the ${\bf 35}$ and ${\bf
405}$ (The $SU(6)$ singlet is absent for $J=1$).  We make this
restatement in order to compare to the nonrelativistic quark model
(NRQM) results, which are obtained using only the ${\bf 35}$.  To see
this, note that the quark magnetic moment operator $(e Q_q/2 M_q)
\otimes \sigma_3$, for arbitrary values of $m_{u,d,s}$, not only fits
into the ${\bf 35}$ rep, but contains as many independent parameters
(three) as the $J=1$ part of the ${\bf 35}$.  The NRQM results when,
in addition, we set $m_u = m_d$, so that the number of independent
parameters reduces to two.

	To illustrate this point, let the three initially independent
parameters in the $J=1$ part of the ${\bf 35}$ be labeled $\mu_1$,
$\mu_Y$, and $\mu_{I_3}$ to indicate their $SU(3)$ content.  In order
to relate these parameters to quark magnetic moments, one must adopt
normalizations consistent with those of the corresponding $SU(3)$
generators:
\begin{eqnarray}
\mu_{1} & = & \mbox{} +\f{k}{\s{3}} (\mu_u + \mu_d + \mu_s), \nonumber
\\ \mu_{Y} & = & \mbox{} +\f{k}{\s{6}} (\mu_u + \mu_d - 2\mu_s),
\nonumber \\ \mu_{I_3} & = & \mbox{} +\f{k}{\s{2}} (\mu_u - \mu_d),
\end{eqnarray}
where $k$ is a proportionality constant that is undetermined, because
group theory alone does not set overall scales.  The constraint $m_u =
m_d$ becomes $\mu_u = -2\mu_d$, or
\begin{equation}
\s{6} \mu_1 + \s{3} \mu_Y - \mu_{I_3} = 0 .
\end{equation}
On the other hand, one may read off directly from our $SU(6)$
Clebsch--Gordan tables:
\begin{equation}
\mu_{p,n} = \mbox{} +\f{1}{18\s{2}} (\s{6} \mu_1 + \s{3} \mu_Y \pm 5
\mu_{I_3}) ,
\end{equation}
and between these two equations one immediately obtains $\mu_p/\mu_n =
-3/2$.

\subsection{Higher Multipole Moments}
	Virtually none of the electric quadrupole or magnetic octupole
moments are experimentally accessible; measured values exist only
for the transition quadrupole moment $Q_{p \Delta^+}$; thus a
numerical analysis of the $SU(6)$ relations would be meaningless.
However, for completeness, we display the quadrupole moment relations.
In this sector there are 18 independent parameters (10 decuplet
moments and 8 octet-decuplet transitions) and 12 parameters associated
with the ${\bf 2695}$ rep. The 12 relations are given by
\begin{eqnarray}
\underline{(SU(3),I)} & \hspace{1em} & \underline{\mbox{Electric
quadrupole moment combination}} \nonumber \\
(8_1,0) & & \mbox{} +[(Q_{\Delta^{+ \! +}} + Q_{\Delta^+} +
Q_{\Delta^0} + Q_{\Delta^-}) - (Q_{\Xi^{*0}} + Q_{\Xi^{*-}}) -
2Q_{\Omega^-}] \nonumber \\ & &
\mbox{} - 2\s{2} [(Q_{\Sigma^+ \Sigma^{*+}} + Q_{\Sigma^0 \Sigma^{*0}}
+ Q_{\Sigma^- \Sigma^{*-}}) + (Q_{\Xi^0 \Xi^{*0}} + Q_{\Xi^-
\Xi^{*-}})], \nonumber \\
(8_1,1) & & \mbox{} +[(3Q_{\Delta^{+ \! +}} + Q_{\Delta^+} -
Q_{\Delta^0} - 3Q_{\Delta^-}) + 2(Q_{\Sigma^{*+}} - Q_{\Sigma^{*-}}) +
(Q_{\Xi^{*0}} - Q_{\Xi^{*-}})] \nonumber \\ & &
\mbox{} - 2\s{2} [2(Q_{p \Delta^+} + Q_{n \Delta^0}) + (Q_{\Sigma^+
\Sigma^{*+}} - Q_{\Sigma^- \Sigma^{*-}}) + \s{3} Q_{\Lambda
\Sigma^{*0}} \nonumber \\ & & \mbox{} \hspace{3em} \mbox{} +
(Q_{\Xi^0 \Xi^{*0}} - Q_{\Xi^- \Xi^{*-}})], \nonumber \\
(10,1),(\overline{10},1) & & \mbox{} + (Q_{p \Delta^+} + Q_{n
\Delta^0}) - (Q_{\Sigma^+ \Sigma^{*+}} - Q_{\Sigma^- \Sigma^{*-}}) -
(Q_{\Xi^0 \Xi^{*0}} - Q_{\Xi^- \Xi^{*-}}), \nonumber \\
(27_1,0) & & \mbox{} + 4[3(Q_{\Delta^{+ \! +}} + Q_{\Delta^+} +
Q_{\Delta^0} + Q_{\Delta^-}) - 5(Q_{\Sigma^{*+}} + Q_{\Sigma^{*0}} +
Q_{\Sigma^{*-}}) \nonumber \\ & & \mbox{} \hspace{2em} \mbox{} -
3(Q_{\Xi^{*0}} + Q_{\Xi^{*-}}) + 9Q_{\Omega^-}] \nonumber \\ & &
\mbox{} + 7\s{2} [2(Q_{\Sigma^+ \Sigma^{*+}} + Q_{\Sigma^0 \Sigma^{*0}}
+ Q_{\Sigma^- \Sigma^{*-}}) - 3(Q_{\Xi^0 \Xi^{*0}} + Q_{\Xi^-
\Xi^{*-}})], \nonumber \\
(27_1,1) & & \mbox{} + 8[(3Q_{\Delta^{+ \! +}} + Q_{\Delta^+} -
Q_{\Delta^0} - 3Q_{\Delta^-}) - 3(Q_{\Sigma^{*+}} - Q_{\Sigma^{*-}}) -
4(Q_{\Xi^{*0}} - Q_{\Xi^{*-}})] \nonumber \\ & &
\mbox{} - 7\s{2} [(Q_{p \Delta^+} + Q_{n \Delta^0}) + 3(Q_{\Sigma^+
\Sigma^{*+}} - Q_{\Sigma^- \Sigma^{*-}}) - 2\s{3} Q_{\Lambda
\Sigma^{*0}} \nonumber \\ & & \mbox{} \hspace{3em} \mbox{} -
2(Q_{\Xi^0 \Xi^{*0}} - Q_{\Xi^- \Xi^{*-}})], \nonumber \\
(27_1,2) & & \mbox{} + 8[3(Q_{\Delta^{+ \! +}} - Q_{\Delta^+} -
Q_{\Delta^0} + Q_{\Delta^-}) + (Q_{\Sigma^{*+}} - 2Q_{\Sigma^{*0}} +
Q_{\Sigma^{*-}})] \nonumber \\ & &
\mbox{} - 7\s{2} [3(Q_{p \Delta^+} - Q_{n \Delta^0}) + (Q_{\Sigma^+
\Sigma^{*+}} - 2Q_{\Sigma^0 \Sigma^{*0}} + Q_{\Sigma^- \Sigma^{*-}})],
\nonumber \\
(35,1),(\overline{35},1) & & \mbox{} + (Q_{p \Delta^+} + Q_{n
\Delta^0}) - (Q_{\Sigma^+ \Sigma^{*+}} - Q_{\Sigma^- \Sigma^{*-}}) -
2\s{3} Q_{\Lambda \Sigma^{*0}} \nonumber \\ & & \mbox{} \hspace{1em}
\mbox{} + 2(Q_{\Xi^0 \Xi^{*0}} -_{\Xi^- \Xi^{*-}}), \nonumber \\
(35,2),(\overline{35},2) & & \mbox{} + (Q_{p \Delta^+} - Q_{n
\Delta^0}) - (Q_{\Sigma^+ \Sigma^{*+}} - 2Q_{\Sigma^0 \Sigma^{*0}} +
Q_{\Sigma^- \Sigma^{*-}}),
\nonumber \\
(64,0) & & \mbox{} + (Q_{\Delta^{+ \! +}} + Q_{\Delta^+} +
Q_{\Delta^0} + Q_{\Delta^-}) - 4(Q_{\Sigma^{*+}} + Q_{\Sigma^{*0}} +
Q_{\Sigma^{*-}}) \nonumber \\ & & \mbox{} \hspace{1em} \mbox{} +
6(Q_{\Xi^{*0}} + Q_{\Xi^{*-}}) - 4Q_{\Omega^-}, \nonumber \\
(64,1) & & \mbox{} + (3Q_{\Delta^{+ \! +}} + Q_{\Delta^+} -
Q_{\Delta^0} - 3Q_{\Delta^-}) - 10(Q_{\Sigma^{*+}} - Q_{\Sigma^{*-}}) +
10(Q_{\Xi^{*0}} - Q_{\Xi^{*-}}), \nonumber \\
(64,2) & & \mbox{} + (Q_{\Delta^{+ \! +}} - Q_{\Delta^+} -
Q_{\Delta^0} + Q_{\Delta^-}) - 2(Q_{\Sigma^{*+}} - 2Q_{\Sigma^{*0}} +
Q_{\Sigma^{*-}}), \nonumber \\
(64,3) & & \mbox{} + Q_{\Delta^{+ \! +}} - 3Q_{\Delta^+} +
3Q_{\Delta^0} - Q_{\Delta^-} .
\end{eqnarray}

	The situation for the octupole moments is in fact trivial.
There are 10 parameters and 10 relations, because the $J=3$ block of
the $92 \times 92$ orthogonal matrix is identical to the pure $SU(3)$
matrix ${\cal C}_b$.  This in turn follows because the only
combinations with $J=3$ originate in decuplet-decuplet bilinears.  The
interpretation of this result is that only the ${\bf 2695}$ rep
contributes to octupole moments, and so these moments, if they are
ever measured, should be numerically uniformly tiny.
\section{Conclusions}
	To summarize our findings, we have shown that one may
conveniently compute all Clebsch--Gordan coefficients associated with
the product $\overline{\bf 56} \otimes {\bf 56}$, and we have
exhibited these coefficients for the particular bilinear combinations
with $\Delta I_3 = \Delta Y = \Delta J_3 = 0$.  All the others, useful
for baryon decay processes, can be obtained from those in this paper
by means of the $SU(2)$ or $SU(3)$ Wigner--Eckart theorem.

	From our coefficients we have compiled all baryon mass and
magnetic moment relations resulting from ignoring the ${\bf 2695}$
component in $SU(6)$.  Violations of the mass relations can be
explained with naive estimates of the neglected operators, and we have
obtained a prediction for the size of $\Sigma^0$-$\Lambda$ mixing.  We
have shown that enough magnetic dipole moments are experimentally
well-known to predict the others, and have used these relations to
show agreement with the experimental value for $\mu_{\Delta^{+ \! +}}$
but disagreement for $\mu_{p \Delta^+}$.  The latter result may be an
indication of the superiority of the large-$N_c$ predictions in
general; the verification of this statement awaits the systematic
analysis of the large-$N_c$ contracted spin-flavor symmetry.
\section*{Acknowledgments}
	I wish to thank Roger Dashen, Elizabeth Jenkins, Aneesh
Manohar, and Mahiko Suzuki for valuable conversations.  This work was
supported by the DOE under grant number DOE-FG03-90ER40546.  It was
begun while I was working at Lawrence Berkeley Laboratory.

\newpage
\setcounter{page}{1}
\thispagestyle{empty}
\begin{table}
\caption{Experimental values for SU(6) mass relations (MeV)}
\begin{tabular}{cc|cc}
(8,0) & \mbox{} $+208.2 \pm 3.5$ & (64,0) & \mbox{} $+5.9 \pm 1.7$ \\
3(27,1) -- (8,1) & \mbox{} $-15.4 \pm 12.7$ & (64,1) & \mbox{} $+0.5
\pm 1.1$ \\ (10,1), ($\overline{10}$,1) & \mbox{} $+0.3 \pm 0.6$ & (64,2) &
\mbox{} $-5.2 \pm 4.5$ \\  (27,0) & \mbox{} $-278.5 \pm 23.2$ &
(64,3) & 0 \\ (27,2) & \mbox{} $+9.1 \pm 5.5$ & &
\end{tabular}
\end{table}
\vspace{5em}
\begin{table}
\caption{Magnetic moment predictions (in $\mu_N$)}
\begin{tabular}{rr|rr|rr}
$\mu_{\Sigma^0}$ & $0.86 \pm 0.30$ & $\mu_{\Sigma^{*0}}$ & $0.37 \pm
0.45$ & $\mu_{\Sigma^+ \Sigma^{*+}}$ & $2.05 \pm 0.04$ \\
$\mu_{\Delta^{+ \!  +}}$ & $5.42 \pm 0.49$ & $\mu_{\Sigma^{*-}}$ &
$-2.94 \pm 0.06$ & $\mu_{\Sigma^0 \Sigma^{*0}}$ & $1.04 \pm 0.21$ \\
$\mu_{\Delta^+}$ & $3.10
\pm 0.46$ & $\mu_{\Xi^{*0}}$ & $0.60 \pm 0.22$ & $\mu_{\Sigma^-
\Sigma^{*-}}$ & $-0.26 \pm 0.04$ \\ $\mu_{\Delta^0}$ & $0.16 \pm 0.45$ &
$\mu_{\Xi^{*-}}$ & $-2.46 \pm 0.23$ & $\mu_{\Lambda \Sigma^{*0}}$
& $2.22 \pm 0.09$ \\ $\mu_{\Delta^-}$ & $-3.41 \pm 0.50$ & $\mu_{p
\Delta^+}$ & $2.52 \pm 0.23$ & $\mu_{\Xi^0 \Xi^{*0}}$ & $2.07 \pm 0.12$ \\
$\mu_{\Sigma^{*+}}$ & $3.05 \pm 0.04$ & $\mu_{n \Delta^0}$ & $2.81 \pm
0.23$ & $\mu_{\Xi^- \Xi^{*-}}$ & $-0.26 \pm 0.12$
\end{tabular}
\end{table}
\end{document}